\documentclass[usenatbib]{mn2e}
\usepackage{amssymb, amsmath}
\usepackage[caption=false]{subfig}
\usepackage{graphicx} 
\title{The Luminous Convolution Model for spiral galaxy rotation curves}
 \author[S. ~Cisneros et. al]{S.\,~Cisneros$^{1}$ \thanks{E-mail:cisneros@mit.edu}, J.\,~G.\,~O'Brien$^{2}$, N.\,~S.~Oblath$^{1}$, and  J.\,~A.~Formaggio$^{1}$\\
 $^{1}$ Department of Physics, Massachusetts Institute of Technology, Cambridge, MA~02139, USA \\
$^{2}$ Department of Sciences, Wentworth Institute of Technology,
Boston, MA~02130, USA  }
 \begin{document}

\maketitle
 
\begin{abstract}
The Luminous Convolution Model (LCM)  is an empirical formula, based on  a heuristic convolution of  Relativistic transformations,  which makes it possible to predict the observed rotation curves of a broad class of spiral galaxies from luminous matter alone.   Since the LCM is independent of distance estimates or dark matter halo densities, it is the first model of its kind which constrains luminous matter modeling directly from the observed spectral shifts of characteristic photon emission/absorption lines.   In this paper we present the LCM solution to a diverse sample of twenty-five (25) galaxies of varying morphologies and sizes.   
For the chosen sample, it is shown that the LCM is more accurate than either Modified Newtonian Dynamics or dark matter models and returns physically reasonable mass to light ratios and exponential scale lengths. 
Unlike either Modified Newtonian Dynamics or dark matter models, the LCM predicts something which is directly falsifiable through improvements in our observational capacity, the luminous mass profile. 
The question, while interesting, of if the LCM constrains the relation of the baryonic to dark matter is beyond the scope of the current work. 
  
The focus of this paper is to show that it is possible to describe a broad and diverse spectrum of galaxies efficiently with the LCM formula.  Moreover, since  the LCM free parameter predicts the ratio of the Milky Way galaxy baryonic mass  density to that of the galaxy emitting the photon, if the Milky Way mass models can be trusted at face values, we then show that the LCM becomes a zero parameter model.   
 This paper substantially expands the results in arXiv:1309.7370 and arXiv:1407:7583.
 \end{abstract}
 \section[]{Introduction}
 
Flat rotation-curve observations of spiral galaxies  have long  been considered  the smoking gun for the existence of dark matter to account for the missing mass problem~\citep{1978Rubin,Bosma78}.    In recent years   novel new techniques have been employed in the search for dark matter.   However at the present moment, without the parameter space being reduced, the lack of  pure evidence  has paved the way for researchers to explore alternative gravitational models.  Many of these theories such as Modified Newtonian Dynamics (MOND)~\citep{Milgrom} and Conformal Gravity \citep{fitting} have shown success and shortcomings.  Here, we present  a new prescription,  the Luminous Convolution Model (LCM),  for calculating galaxy rotation curves using  a modified velocity addition formula with  only luminous matter.  The LCM is consistent with Special and General Relativity (SR+GR), though uses  new applications  of these familiar concepts.  \\
Traditionally, the relative curvature effects, i.e. gravitational redshifts  of a  photon from the  spiral  galaxy where it is  emitted   with respect  to  where it is received,  are neglected as   too small to impact the total rotation curve magnitudes. However, such relative curvatures effects are evaluated by taking the algebraic difference of the gravitational redshifts of the two galaxies in question.   We posit that this  classical subtraction of gravitational redshifts is a Galilean concept, and therefore not applicable to light.  Since light remains  invariant under Lorentz transformations, not Galilean,  we re-calculate the relative curvature effects from luminous matter using Lorentz-type transformations in order to relate the  relative galaxy frames.  The   equivalent doppler shifts for the emitter galaxy relative to the receiver galaxy (the Milky Way) are then re-phrased   kinematically, and we show how this new effect can account for the missing    rotational velocity in an arbitrary spiral galaxy.  \\

 In a sample of  twenty-five $(25)$   well studied  galaxies  
 the Luminous Convolution Model  (LCM)    fits rotation curves more accurately than either Modified Newtonian Dynamics  (MOND) or dark matter halo (DM) fits.  
 The LCM is    an empirical  formula which     predicts  the observations of spiral galaxy rotation curves   with reasonable estimates of the luminous mass,   across the samples broad range of  sizes and morphologies without modifying classical laws of gravity.  Additionally, the LCM  is a potentially falsifiable model, through comparisons between observations and the density predictions of the luminous mass made from the LCM free parameter. 
In this paper,  our   analysis is  focused   on   the highly   symmetric  case of spiral galaxies (in the plane of the galactic disk), but  it should be noted that the theoretical  LCM mapping basis  can   in theory be  analytically  extended to  arbitrary  geometries~\citep{Cisneros:2013vha}.  Examples  of physical systems where LCM constructions  will  be considered in future work,    which  are  beyond the scope of the current paper, include
 galaxy mergers, clusters of galaxies, lensing, and elliptical galaxies. 
 
 This paper is organized as follows: Sec.~\ref{sec:DERIVE} is the summary description of the LCM mapping formalism,  Sec.~\ref{sample} presents the sample and results, Sec.~\ref{sec:conclusion} presents  conclusions and possible future directions and Sec.~\ref{sec:fin}  gives acknowledgements and an appendix with a detailed LCM heuristic derivation.
 
  \section[]{Lorentz kinematics and the LCM rotation curve formula }
\label{sec:DERIVE} 
 
 To account for the missing mass in spiral galaxies, DM theories contain a contribution of rotational velocity  from the dark masses, resulting in the rotation curve formula,
  \begin{equation}
v_{rot}^2 =  v_{lum}^2 +  v_{dark}^2,
\label{eq:zonte1}
\end{equation} 
where $v_{lum}$ is the contribution expected from photometric observations of the luminous matter (disk, bulge and gas).  The typical functional form of the expected value of $v_{lum}$  is the Freeman formula,
\begin{equation}
v^2_{lum} = \frac{N^*\beta^*c^2R^2}{2R^3_0}F( R )  \label{gr}
\end{equation}
where {$R_o$} is the galactic scale length and {$N^*$} is the number of stars in the galaxy.  The function F is given by
\begin{equation}
F( R )=\left[I_0\left(\frac{R}{2R_0}\right)K_0\left(\frac{R}{2R_0}\right)-I_1\left(\frac{R}{2R_0}\right)K_1\left(\frac{R}{2R_0}\right)\right],
\end{equation}
where $I_0$, $I_1$, $K_0$, and $K_1$ are the standard modified Bessel functions. Although the dark matter contributions are not detectable by any current means, the relative amounts in each galaxy are inferred by the difference between   the  expected   luminous mass contribution and  the total rotation observations, $v_{obs}$, which comes from measurements of Doppler shifted spectra.  Hence the sum  $v_{rot}$ is the DM   prediction which is fit against the actual measured velocity $v_{obs}$.
  
  The functional forms of the two terms in Eq.~(\ref{eq:zonte1})  have been well established (see for example \citep{mfull}) and usually provide an accurate statistical fit to the data.  However, since the luminous contribution can only typically account for the inner region of spiral galaxies, the dark component dominates the outer regions, and fits the data with two free parameters per galaxy.
 
 In our prescription, the Lorentz Doppler shift formula  (LD) 
  \begin{equation}
 \frac{v_{obs}(r)}{c}=
\frac{  \frac{\omega'(r)}{\omega_o}- \frac{\omega_o}{\omega'(r)}}{ 
\frac{\omega'(r)}{\omega_o}+ \frac{\omega_o}{\omega'(r)}},
\label{eq:dataLorentz}
\end{equation}
which  acts to convolve the  frequency  $\omega'$ (received from a moving frame)  with the characteristic rest frame frequency $\omega_o$.  Then $\omega_o$ is generalized so that 
the resulting velocity parameter $v_{obs}$  in Eq.~(\ref{eq:dataLorentz}) is interpreted  as the underlying Lorentz transformation which rotates a photon's four-vector between the  two frames.   It is from this formula  which 
 we will derive a luminous photometric profile that will match  $v_{obs}$, by rotating between the 
 two slightly curved galactic frames of spiral galaxies, based on decomposing the total  
   observed Doppler-shifted frequencies  $\omega'$ into  two contributions: the relative velocity and the relative curvature.

Hence, the LCM modifies the  velocity addition formula in  Eq.~(\ref{eq:zonte1}) by replacing the 
DM halo velocity contributions, $v^2_{dark}$, with the relative curvature  convolution $\tilde{v}^2_{lcm}$ .  It has been shown in previous work ~\citep{Cisn,Radosz} that curvature effects can be phrased kinematically.  The LCM prediction of the total  
  observed shifted frequencies, $\omega'$ interpreted via the LD,  is then :
  \begin{equation}
v_{rot}^2 =  v_{lum}^2+ \alpha \tilde{v}_{lcm}^2 ,
\label{eq:zonteLCM}
\end{equation}  
where $ v_{lum}^2$ remains the  expected relative velocity  (Freeman) contribution,   $v_{rot}$ is   the LCM prediction which is fitted to the total reported $v_{obs}$, and $\alpha$ is the LCM  free  fitting parameter.  Although in the original work on the LCM ~\citep{Cisneros:2013vha},  {$\alpha$} was a purely free parameter, in this work we will show that instead it is   highly correlated to the dimensionless ratio
\begin{equation}
\alpha=\left(\frac{\rho_{mw}}{\rho_{gal}}\right)^{1.64}
\label{correl}
\end{equation}
 of the  radial densities of   the photon    receiving galaxy (Milky Way)  $\rho_{mw}$  to  the photon emitting  galaxy  $\rho_{gal}$.   Here, the radial density is 
 defined by:
 \begin{equation}
 \rho=\frac{M_{total}}{r_e},
 \label{radDens}
 \end{equation}
for $M_{total}$ the  integrated total  luminous mass at the limit of the reported data,  and $r_e$ the exponential scale-length (described in Sec.~\ref{fitting}) for that baryonic mass distribution.   
 
\subsection{ LCM relative curvature convolution}
Relative curvatures for the spiral galaxies presented in this work are defined as a function of 
radius $r$ from the center of the galaxy, 
with the  Schwarzschild metric   gravitational redshifts via:
  \begin{equation}
 \frac{\omega_o }{ \omega ( r) }=\left(\frac{1}{\sqrt{-g_{tt}} }\right)_r.
\label{eq:Clone}
\end{equation}
Here,  $\omega_o$ is  the   characteristic photon frequency (defined in Eq.~\ref{eq:dataLorentz}), and $\omega (r)$ is  the  shifted frequency due to the gravitational curvature\footnote{ as 
  received by a stationary observer  at asymptotic infinity}.  The temporal  Schwarzschild metric   coefficient  $g_{tt}$ is defined as the usual:
 \begin{equation}
 g_{tt}(r)=-\left(1- 2\frac{G M}{c^2r}\right), 
 \label{eq:timeportion}
 \end{equation}
where $G$ is  Newton's constant of gravity,   $M$ is  the  Gaussian enclosed  mass at some  radial distance $r$  from the center of 
 the mass distribution,  and
 $c$ is  the vacuum light speed.     In the weak field limit ~\citep{Hartle}, the temporal metric Eq (\ref{eq:timeportion})  takes the form of
 \begin{equation}
g_{tt}(r) \approx -1 + 2\frac{\Phi(r)}{c^2}.
\label{eq:weakfield}
\end{equation}
Here,     $\Phi$ is the usual Newtonian  scalar   gravitational potential. Spiral galaxies as a whole can be  treated gravitationally as weak fields due to  the diffuse nature of the luminous mass distributions.  This has   served to motivate the common use of the classical Poisson equation in DM theories.  For the LCM, this serves to  justify the use of the Lorentz transformation architecture to  map  the slightly curved frames  of spiral galaxies.  
 
 While the relative curvature estimates we use are precisely those which have previously been obviated as too 
 small to impact rotation curve velocities,  
  we should note the difference between the LCM and previous treatments which have ruled out curvature contributions from luminous matter.  Namely,   previous treatments considered relative curvature effects by simply taking the difference of the magnitudes of the two gravitational redshifts between the photon source (the emitting galaxies) and the Milky Way (receiving galaxy).  This over-simplified explanation is one that is Galilean in nature and will in turn be inadequate.  This can be easily rectified since light is a Lorentz invariant and so requires a relativistic transformation to relate any relative effects that could arise from  curvature, even if they are weak.  
  
 The entire LCM heuristic mapping derivation is described in Appendix A, but here for continuity we present  the resulting $\tilde{ v}_{lcm}^2$:
\begin{equation}
\tilde{ v}_{lcm}^2=   \kappa^2  v_{1}  v_{2},
\label{eq:convolutionFunc}
\end{equation}
where $\kappa$   is the curvature ratio which  scales 
 the coordinate time of  the emitter galaxy relative to that of the  Milky Way:
\begin{equation}
\kappa(r)=\frac{c-\tilde{c}_{gal}(r)}{c-\tilde{c}_{mw}(r)},
\label{eq:kappa}
\end{equation}
 by multiplying each factor of  $c$ which  results from the successive $v_1/c$ and $v_2/c$ Lorentz-type transformations.  The terms 
   $\tilde{c}_{gal} (r)$  and  $\tilde{c}_{mw} (r)$ are the  respective   coordinate light speeds of the emitter   and receiver galaxies.    These coordinate light speeds  are  a physical indicator   of curvature, specifically  the degree to which path lengths are increased due to curvature.    We then define $\tilde{c}$ in terms of the Schwarzschild metric relation:
    \begin{equation}
 n(r) \tilde{c }=\left(\frac{1}{\sqrt{-g_{tt}} }\right)_r  \tilde{c}=c.
 \label{eq:index}
\end{equation}
 
   The  Lorentz-type transformation  $v_1$ acts to map the two galactic frames as a function of radius:
    
      \begin{equation}
\frac{v_{1}}{c}=\left(\frac{1}{\cosh \xi_{c} } -1\right)=\left(\frac{ 2}{e^{\xi_{c}}+ e^{-\xi_{c}}} -1\right),
\end{equation}
  for the fundamental mapping kernel:
   \begin{equation}
e^{\xi_{c}}(r)=\frac{\omega_{mw}(r)}{\omega_{gal}(r)}.
\label{eq:11}
\end{equation}
The respective frequencies $\omega_{mw}$ and $\omega_{gal}$ are the gravitational 
redshift frequencies defined by $\omega (r)$ in Eq.~\ref{eq:Clone} as a function of $r$. 

The second Lorentz-type transformation $v_{2}$ maps from the curved two frame of  Eq.~\ref{eq:11} to  the 
the associated flat frames where photons are emitted and received:
  \begin{equation}
\frac{v_2 }{c}= 1/\tanh \xi_2= \frac{(e^{ \xi_2} )^2+1}{(e^{ \xi_2} )^2 - 1}, 
\label{eq:hyperbolico}
\end{equation}
  for the fundamental mapping kernel:
\begin{equation}
 (e^{ \xi_2} )^2= \frac{e^{\xi_{f}}}{ e^{\xi_{c}}  }.
\end{equation} 
Lastly, the flat $2-$frame mapping is defined by
 \begin{equation}
  e^{\xi_{f}}(r) =   \frac{\omega_{l} (r)}{\omega_{o}}, 
     \end{equation}
which allows for our contribution $\tilde{ v}_{lcm}^2$ to be calculated and then used in Eq.~\ref{eq:zonteLCM} to give the entire LCM prediction for the rotational velocity of the galaxy in question.\\
 
 \subsubsection{Note on integration constants and physicallity: }  
The  gravitational potential,  $\Phi(r)$, which   parametrizes the curvatures of interest in    Eq.~\ref{eq:weakfield}, is defined as an integral over   the Newtonian force  $F(r)$:
\begin{equation}
\Phi(r)-\Phi_o=-\int \frac{F(r) }{m} dr,
\label{eq:potentialgeneral}
\end{equation}  
where   $F(r) /m$ is the force per unit mass for  each individual  luminous mass component,  and  $\Phi_o$  is the integration constant of interest.

  This   integration constant is  generally set such that the potential $\Phi(r)\to 0$ as $r\to \infty$.  However, when considering     two arbitrary galaxies\footnote{the  emitter galaxy and the receiver galaxy (Milky Way)}, connected by a single photon,  it is a violation of energy conservation to   set   the respective integration constants to different values.   We select a single universal value for the integration constant,  taken to be zero.  It is beyond the scope of this paper to find a more accurate choice, but it should be considered that  a more physical     choice  may be found in   future dark energy  research.   Physically, our assumption means that at large $r$    the gravitational potentials return the familiar small  but non-zero  values.   
 
\subsection{LCM rotation curve fitting protocol}
 \label{fitting}
\begin{figure}
\includegraphics[scale=0.65]{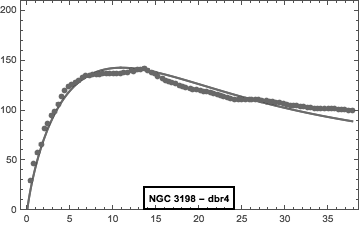} 
\caption{The fit returned in steps (iii) through (v) of the LCM fitting procedure for NGC 3198.  Galaxy name appended with dbr4 indicates original luminous mass model from  \citet{Blok} and the fourth run against the data, where each run iterates the luminous mass profile to reflect the observed spectral line shifts in the reported data. \label{3198}}
\end{figure}

With the basis for the LCM contributions established,  we can then fit the predicted LCM rotation curve velocity $v_{rot}$ (Eq.\ref{eq:zonteLCM}) to   the reported rotation curve data $v_{obs}$ (Eq.~\ref{eq:dataLorentz}).  The LCM prediction  and fit    are calculated using the MINUIT minimization software as implemented in the ROOT data-analysis package~\citep{ROOT}. 
The fitting procedure can be summarized as follows:
\begin{enumerate}
 \item The   luminous mass components reported in each reference (gas, disk, bulge) are   digitized using the software package Graph Click \citep{Graphclick};
 \item   The associated    Newtonian gravitational potential $\Phi$ is calculated for each component and the components are  summed (Eq~\ref{eq:potentialgeneral});
  \item The Schwarzschild metric is then parametrized by the luminous mass distribution as a function of radius, and the  convolution function,  $v_{lcm}$,  is calculated in comparison to the Milky Way luminous mass profile;
  
  \item The  minimization procedure  explores the parameter space to find an optimal luminous  profile    ($v_{obs}$) from the data  ($v_{rot}$).    This process can be iterated   within the bounds of reported     mass-to-light ratios,  distances and  scale lengths.     
     \item The resulting   best-fit  values for luminous mass profile (indicated by the rotation curve data) is then fit to return  an exponential scale length $r_e$ for a smoothed profile including gas, disk and bulge, and the total mass of the emitter galaxy.  A sample of the output is shown in Fig. \ref{3198} ; 
     \item The correlation over the entire sample of the $\alpha$ parameter to the density ratio in Eq.~\ref{radDens} is then used to ascertain the global value of the receiving (Milky Way) galaxy's scale length $r_e$ as reported in   Table~\ref{MWlum} for five different Milky Way  luminous mass models.
\end{enumerate}

It is important to note  that a unique feature of the LCM is contained in steps (iv) and (v) above.  Unlike most alternative dark matter models, we use the LCM formalism to predict a luminous profile for the given galaxy.  Hence, when the LCM returns the convolution function, it is fit to the observed data and then we extract a scale length and a luminous mass in step (v).  These values are  reported in Table ~\ref{sumRESULTS}.  These values act as a prediction for the galaxy in question and then can be compared with established results from photometry and population synthesis modeling.  This feature sets the LCM apart,  since we do not use the scale length or mass as input for the galaxy, and thus have a testable result for each galaxy fit by the LCM.  As mentioned above in step (iv),  in this current work we fit the entire returned luminous profile of  the  galaxy with a thin luminous disk   to return estimates of the total luminous mass and exponential scalelength $r_e$.  Although some of the galaxies  may contain a documented bulge, as noted in Table ~\ref{sumRESULTS}  by an asterisk (*),  the main differentiating feature between observation and Keplerian prediction occurs in regions outside the bulge (usually after the peak velocity which typically occurs at $r=2.2r_e$).  Since most bulge contributions fall off quickly,  in general $r_b<r_e$,  then for this work as a test of concept, we kept the fits described in step (iv) (to the Luminous mass profile resulting from step (iii))  simple while still obtaining physically acceptable results.  Future testing will require more rigor of implementing the resulting profiles and testing the full bulge and HI data of each of the galaxies.  Since the LCM requires information about the emitter galaxy as well as the receiver, then it should be noted that the LCM is not only sensitive to the data used of the emitter galaxy, but also the data used for our own milky way.   To avoid bias as well as make the LCM more robust, we have run the sample of emitter galaxies against five different, well studied, Milky Way luminous mass profiles.  

As expected, the luminous mass profiles resulting from step (iv) above  shows  variation  pending  the choice of the Milky Way luminous profile used in the mappings.  However, it should be noted that this is not a pure fitting alone.  The fits returned come with physical quantities such as the luminous mass and the galactic scale length which can be compared with astronomical data.  The fit to NGC 3198 shown in Fig. \ref{3198} is quite a remarkable fit due to the fitting procedure, but more importantly is physically acceptable.  The fit shown returns both a luminous mass and a scale length which is consistent with documented sources \citep{Blok}, and is recovered from Eq. \ref{eq:zonteLCM} without the need for any dark matter.  Moreover,  since estimates of the Milky  Way's baryonic content are notoriously difficult to determine, this demonstrates that the LCM  can be used to distinguish between different estimates of the Milky Way's luminous mass for   a given sample while all other factors are held constant (ie. extent of reported coverage, etc), see Sec.~\ref{MW}.   Moreover, this feature can be leveraged if the data from the emitter galaxies is accurate and the LCM is empowered,  then results over a reliable sample of galaxies can be used to infer validity of Milky Way data.  Since the LCM requires information about the emitter and receiver galaxy,  the returned luminous profile  in steps (v) and (vi) above
 are  specific to the reported data, at the assumed distance. If either variable is changed, the luminous mass must therefore compensate accordingly.  
 
\section{Results} \label{sample}
 \subsection{The sample}

 The LCM sample reported in this paper  represents twenty-five $(25)$   galaxies selected to represent a broad spectrum of morphologies and luminous mass density profiles.  It can be seen in Table ~\ref{sumRESULTS}  that the sample also encompasses a large variation in estimated distances as well as overall size of the galaxies.  Having such a diverse sample allows the LCM to illustrate its modeling power.  Typically in DM theories the morphology of the galaxy (such as dwarf or large spiral) dictates the relative location of the DM and thus changes the corresponding function which fits the data.  This occurs also in competing alternative theories such as MOND, where the choice of the interpolation function can be dictated by properties such as the morphology or the distance.  Here we show that the LCM can account for the missing mass discrepancy in each of the galaxies in a more universal manner.   It should also be noted that the chosen sample is comprised of some of the most r well studied galaxies in rotation curve physics, making the sample as unbiased and reliable as possible~\citep{salucci}.  We also report in Table ~\ref{sumRESULTS}  the original sources of the rotation curve data for more specific information on a particular galaxy. 
 
   Results for each emitter galaxy are given in Table~\ref{sumRESULTS}, including:
    $\alpha$,  M/L, $r_e$,  the LCM    reduced $\chi^2$,  the  reduced $\chi^2$ values for the DM or  alternative gravity model which was originally  fit  to the same data.    The   M/L  reported here reflect the  smoothed profile including the gas, bulge and disk together, as noted in Sec.~\ref{fitting}.   All result values and rotation curve figures (Fig.~ \ref{fig:results3} - \ref{galaxiesS}) reported here reflect the     mapping to a  Milky Way  luminous mass model synthesized from the   \citet{Xue}  disk and \citet{Sofue} bulge.   More information on the Milky Way  data sets  are described in Sec.~\ref{MW}.    
  As can be seen in Fig.~ \ref{fig:results3} - \ref{galaxiesS}, the relative curvature contribution serves to correct the rotation curve and provide outstanding fits. Furthermore since the relative curvature contribution is unique to the Milky Way,  and is not described by any free parameters, it makes the fits shown in Fig. 3-5 very accurate and falsifiable because unlike MOND and dark matter models, the LCM predicts something which we can measure, the baryonic mass. 
 \subsection{Adopted Distances}
 It is important when fitting rotation curves to any theory that a solid understanding of the estimated distances to the galaxy is provided.  As mentioned in the previous section, many alternative theories are quite sensitive to distance modifications to the galactic data since photometric properties such as scale length and luminosity are completely dependent on the assumed distance when the data is taken by an astronomer.  In order to keep the systematics of this work standardized, we will adopt the most common value for the distances to each galaxy captured in the literature.  For most of the sample, this means that we are keeping the original distances quoted by the astronomers, with the exception of some of the Ursa Major Galaxies whose distances have been significantly modified in recent work \citep{SanMcGa}.  We note that the LCM, like most other theories is sensitive to the distance used.  However, in this work we have shown in Table~\ref{sumRESULTS} that using the  quoted distances the LCM is able to return physically reliable scalelengths and mass to light ratios. \cite{fitting} provides an extensive discussion on how to adjust rotation curve data via updated observable parameters such as distances and inclinations, where they use a standard of adopting only distances based on either cepheid data or an averaging over the NASA Extragalactic Database (NED).  In this work we kept true to the quoted distances to show that physical solutions are possible in the LCM, but it should be noted that updated distances are obtained, all the quoted parameters in Table~\ref{sumRESULTS} will scale accordingly.\\

 \subsection{Error estimates and  reduced $\chi^2_r$ values}
All figures reported here indicate the uncertainties reported in the literature.
There is currently no  standard practice  as to how to quantify the uncertainties associated with   rotation curve data  ~\citep{Gent,Blok,JNav,San96},  such that   $\chi^2_r$ values do not indicate global goodness of fit between data sets, but can be used to distinguish models applied to the same data set.  Generally the uncertainties in this paper come  from either      statistical errors from  tilted ring model fits to the  H\,{\sevensize\bf I} velocity fields or  differences between the approaching- and receding-side velocity fields~\citep{Blok,Gent,Toky} .  
In  Table~\ref{sumRESULTS} we show that     $\chi^2_r$ values are consistently lower for LCM fits than for   those of the reporting models.   We use the same comparison of $\chi^2_r$, though averaged across the entire sample to compare different Milky Way luminous mass models, as indicated in Sec.~\ref{MW}.

 \subsection{Milky Way luminous mass models and identification of the $\alpha$ parameter}
\label{MW}  
 
In the LCM  we map each emitter galaxy onto  the receiver galaxy (i.e. the Milky Way) to derive the relative curvature contribution,  $v_{lcm}$,  to the total measured rotation curve.   Since we   are making observations of photons  from the inside of the system,
the luminous matter profile of the Milky Way is  difficult to determine and constrain  (e.g.  interpreting  H\,{\sevensize\bf I} ~\cite{Car}).     We have   compared each emitter galaxy in our sample of twenty-five (25) galaxies to four $(4)$ different    Milky Way   luminous mass models,  to accommodate   these differences in  Milky Way (MW)  luminous mass estimates; \citet{Sofue}, ~\citet{Xue} and two models from \citet{Klypin}.   Table~\ref{tab:MWlum} and Fig.~\ref{MWlum} show the differences between the various  MW  models graphically and numerically.

As can be seen in Table~\ref{tab:MWlum}, these MW models differ primarily in the inner mass contributions and generally asymptote to similar values at large radii.  We have found that in the cases of the Klypin and Xue MW models, the lower inner velocity profile preclude   fitting 
very dense central mass distributions in galaxies such as NGC 2841, NGC 7814, NGC 7331 and NGC 5055.  For the Xue model,  the $\alpha$ parameter
 can be constrained to return reasonable fits to  all of these galaxies.   However, it is the Sofue MW which fits all galaxies in our sample with no artificial bounding on the free parameter.    Since one advantage of the Xue MW is the extent of the radial coverage, we have synthesized a MW from the Sofue bulge and the Xue disk, which allows us to fit all galaxies in our sample to their furthest extent, while still taking advantage of the central mass concentration reported in the Sofue MW. 
Since one of the goals  of the current paper is to identify the physical interpretation of the LCM free parameter  $\alpha$, then fitting the galaxies to various MW models helps us to constrain and eventually fix $\alpha$.
 
\begin{figure*}
 \centering
\subfloat[][Xue/Sofue synthesis]{\includegraphics[width=0.33\textwidth]{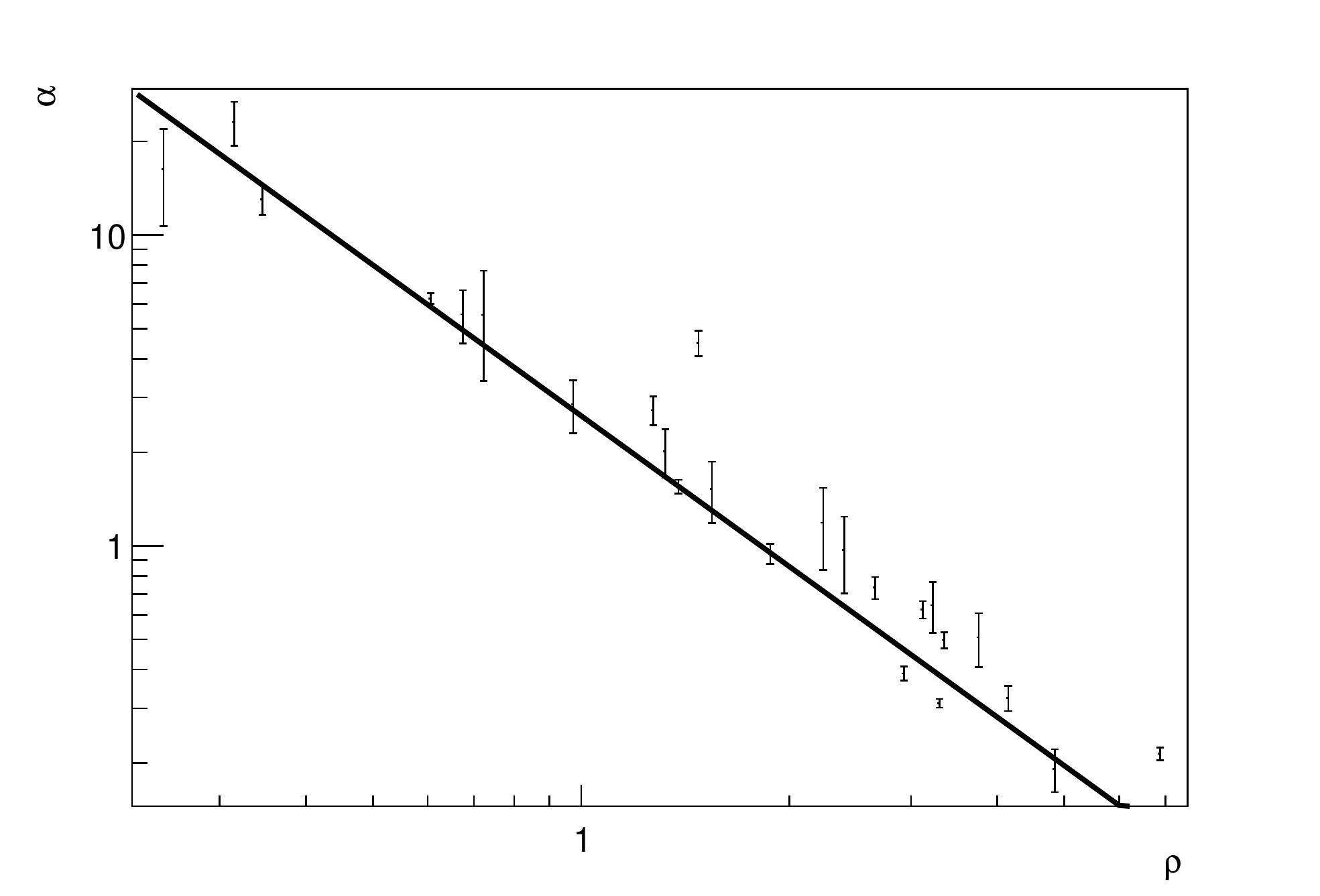}}
\subfloat[][Sofue]{\includegraphics[width=0.33\textwidth]{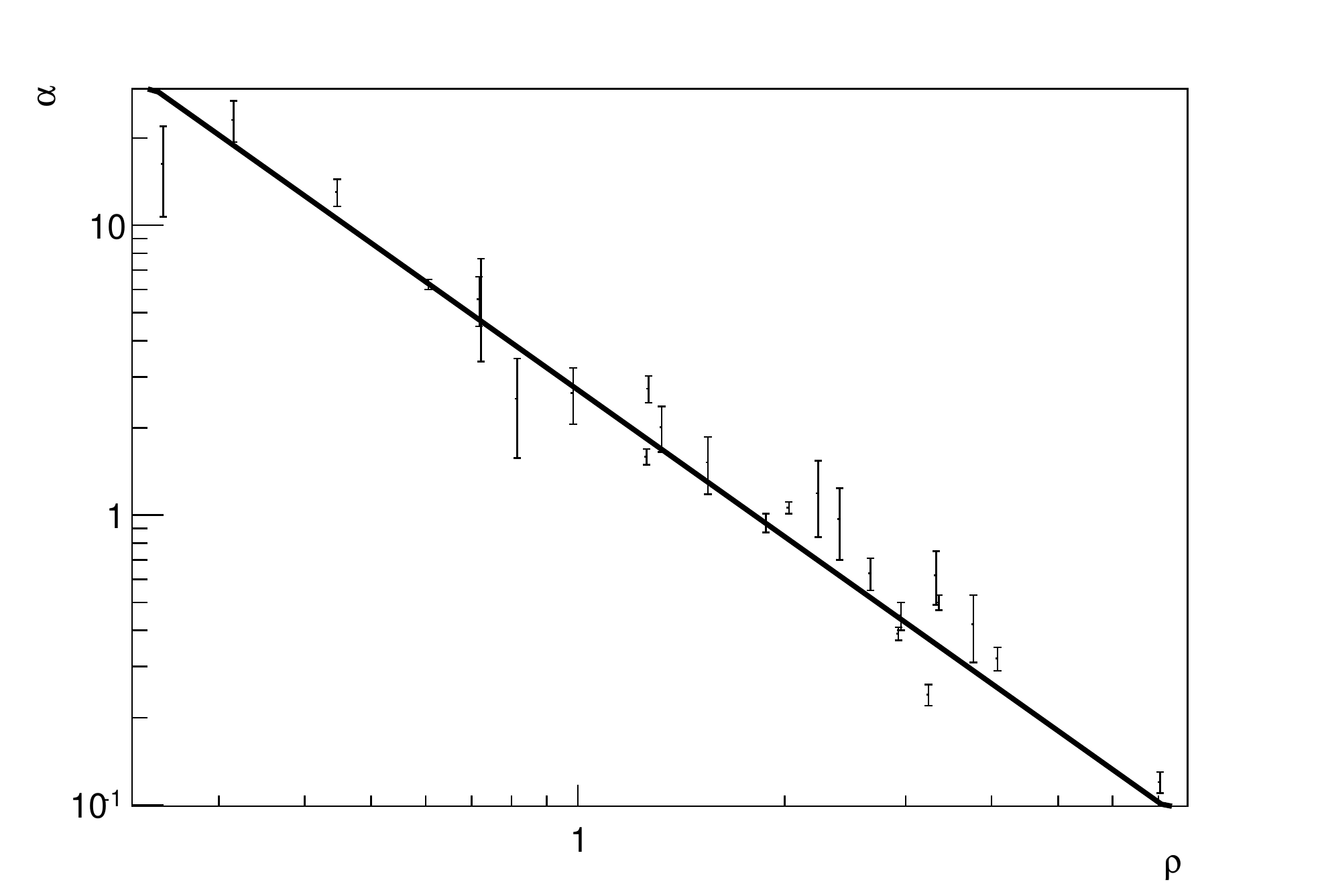}}
\subfloat[][ Xue ]{\includegraphics[width=0.33\textwidth]{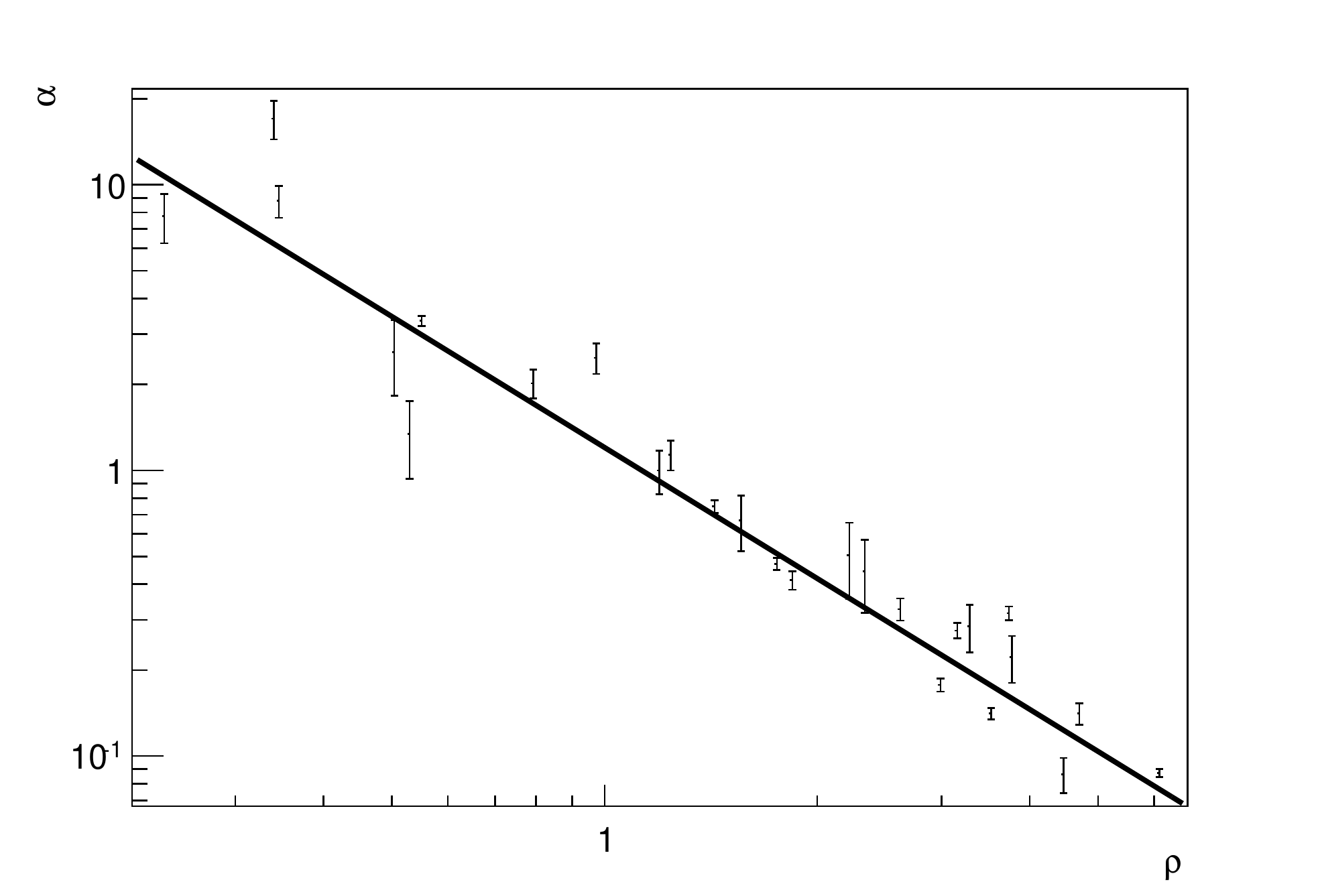}}
\\
\vspace{0.5cm} 
\subfloat[][ Klypin model A]{\includegraphics[width=0.33\textwidth]{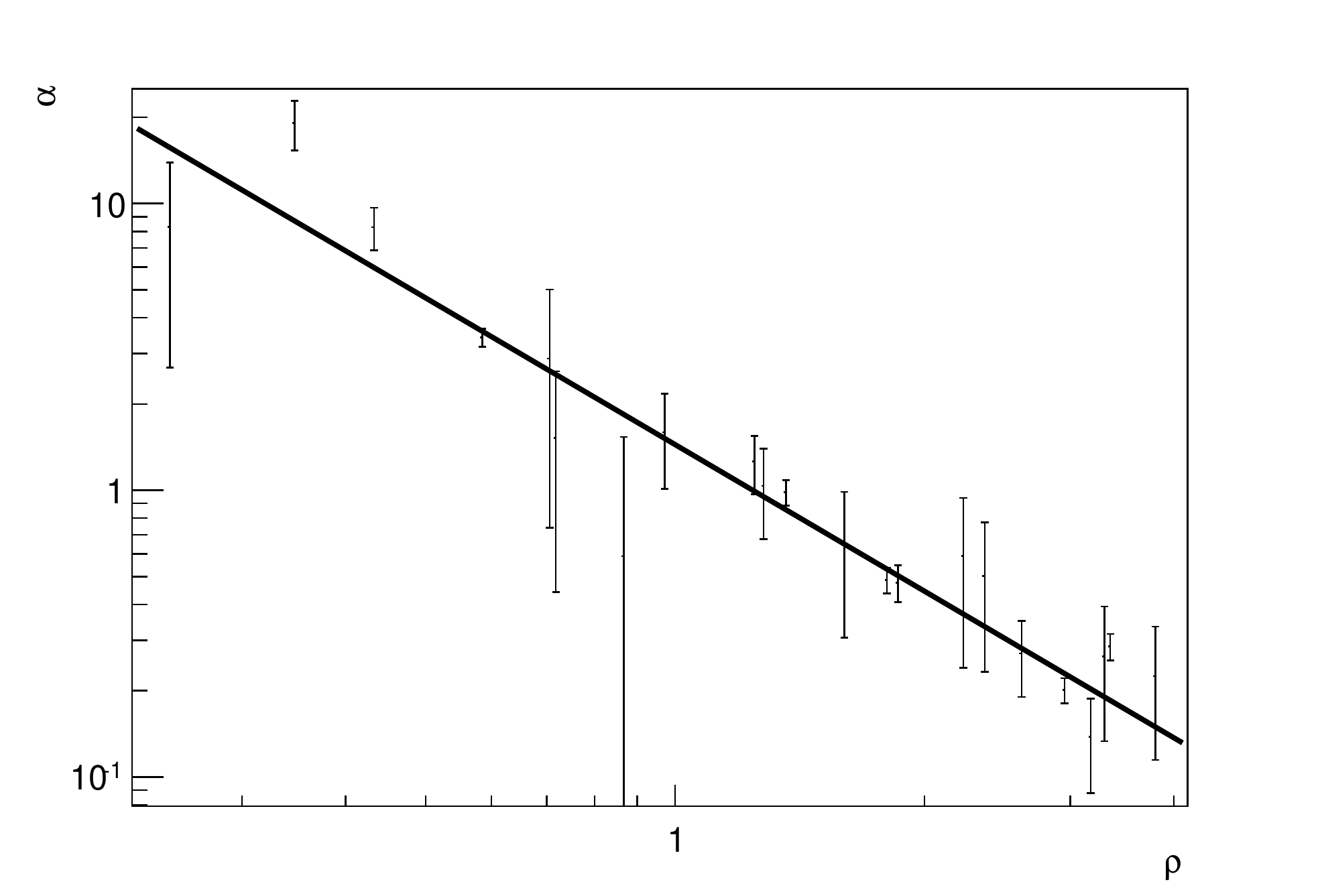}}
\subfloat[][ Klypin model B ]{\includegraphics[width=0.33\textwidth]{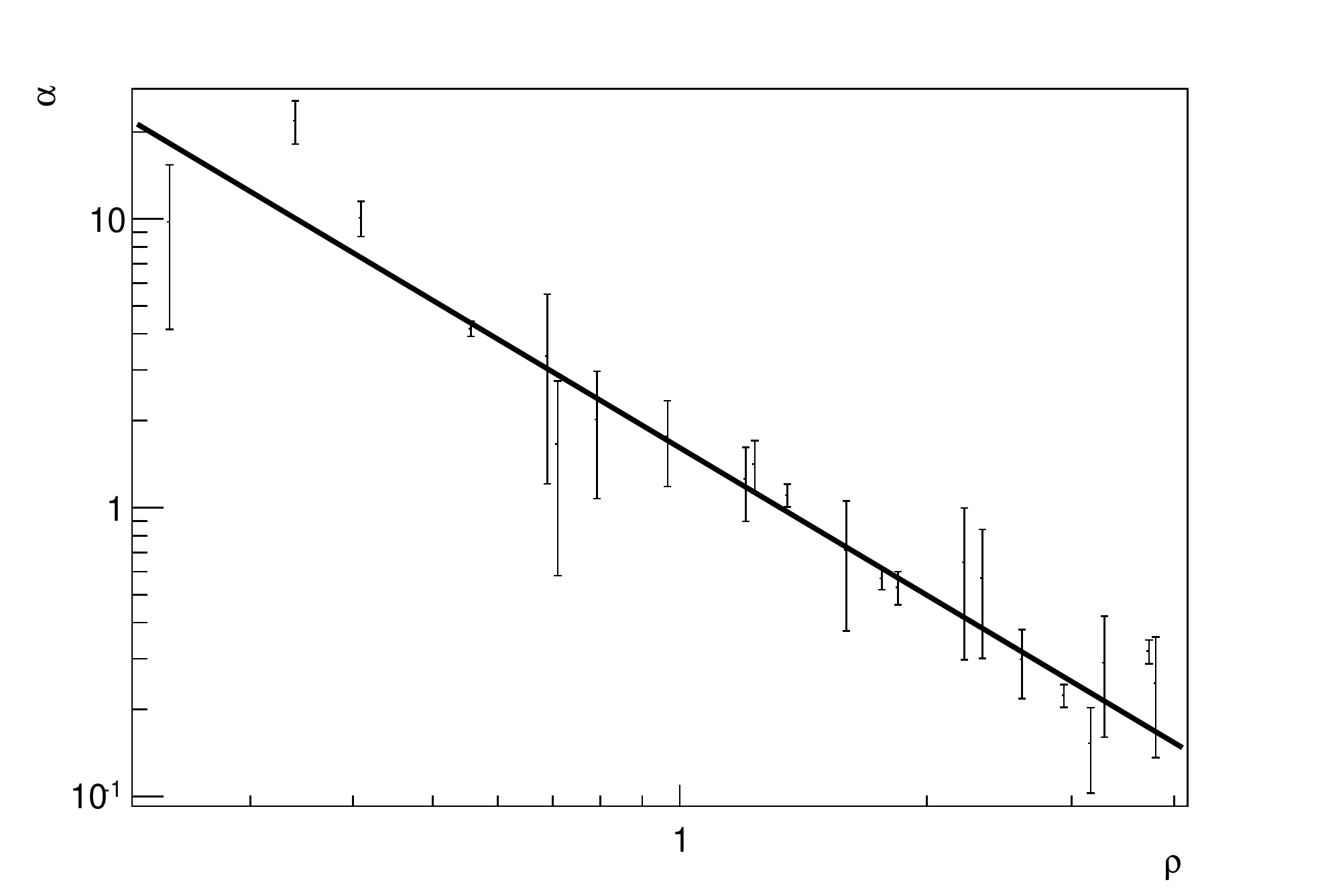}} 
 
 \caption{Log-log plot of the LCM distribution of  $\alpha$   versus  radial density    $\rho_r $ ( Eq.~\ref{radDens}) for one Milky Way   to the sample of photon emitter galaxies,   each dot represents one such galaxy mapped to the Xue/Sofue synthesized Milky Way.  Errors  are statistical only. }    
           \label{fig:alphaDistrib1}
\end{figure*}

\begin{figure}
\includegraphics[scale=0.45]{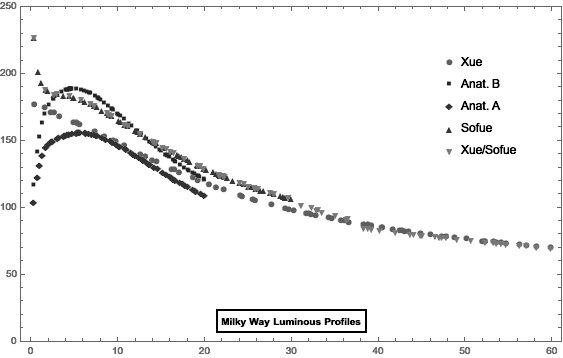} 
\caption{The Milky Way orbital velocities due to the luminous mass  profiles reported in each of the four references used in this work, as well as the Milky Way synthesized from Xue/Sofue.   \label{MWlum}}
\end{figure}

\begin{figure*}
 \centering
\subfloat[][M 31,  Ref.~12]{\includegraphics[width=0.33\textwidth]{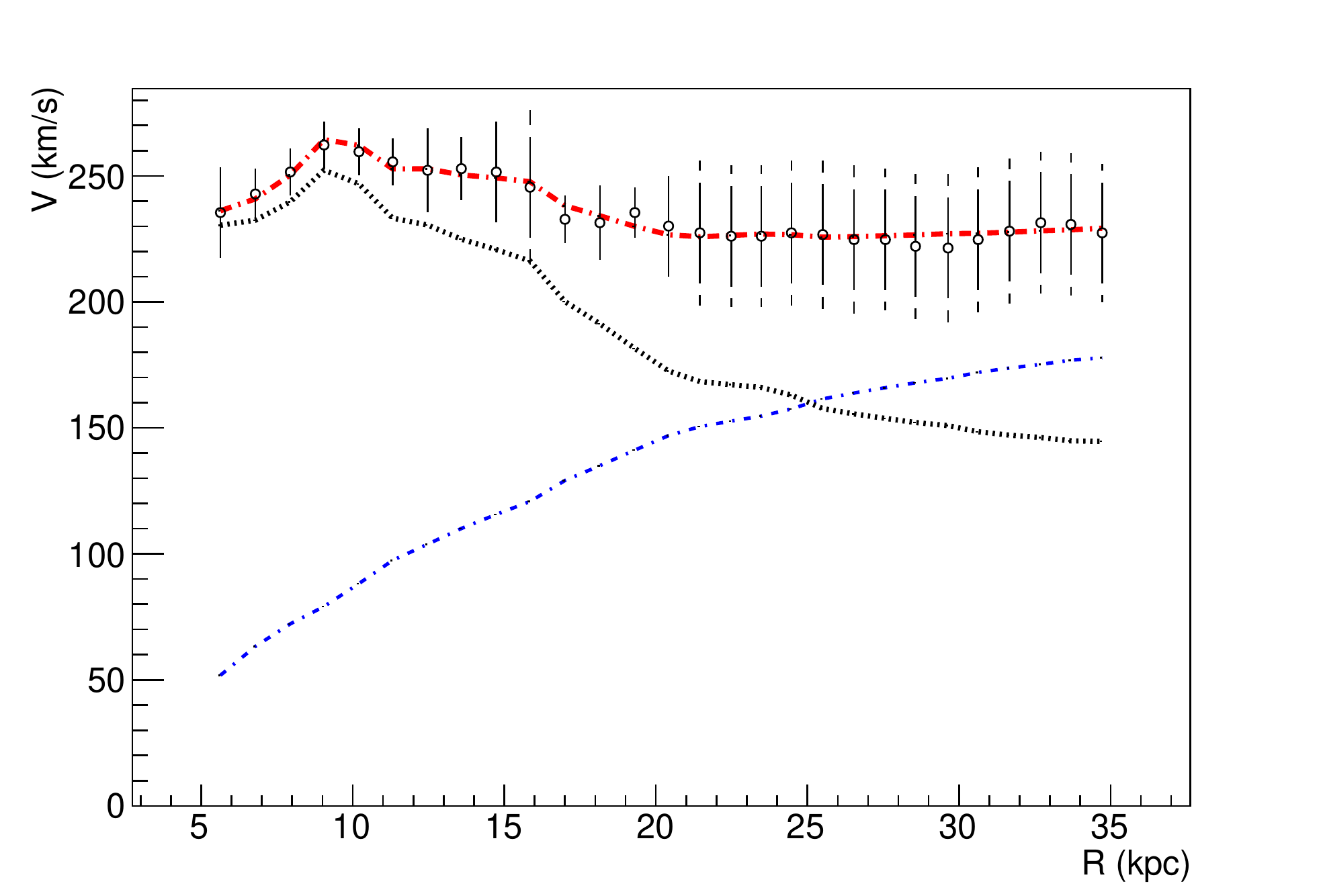}}
\subfloat[][NGC 5533, Ref.~10]{\includegraphics[width=0.33\textwidth]{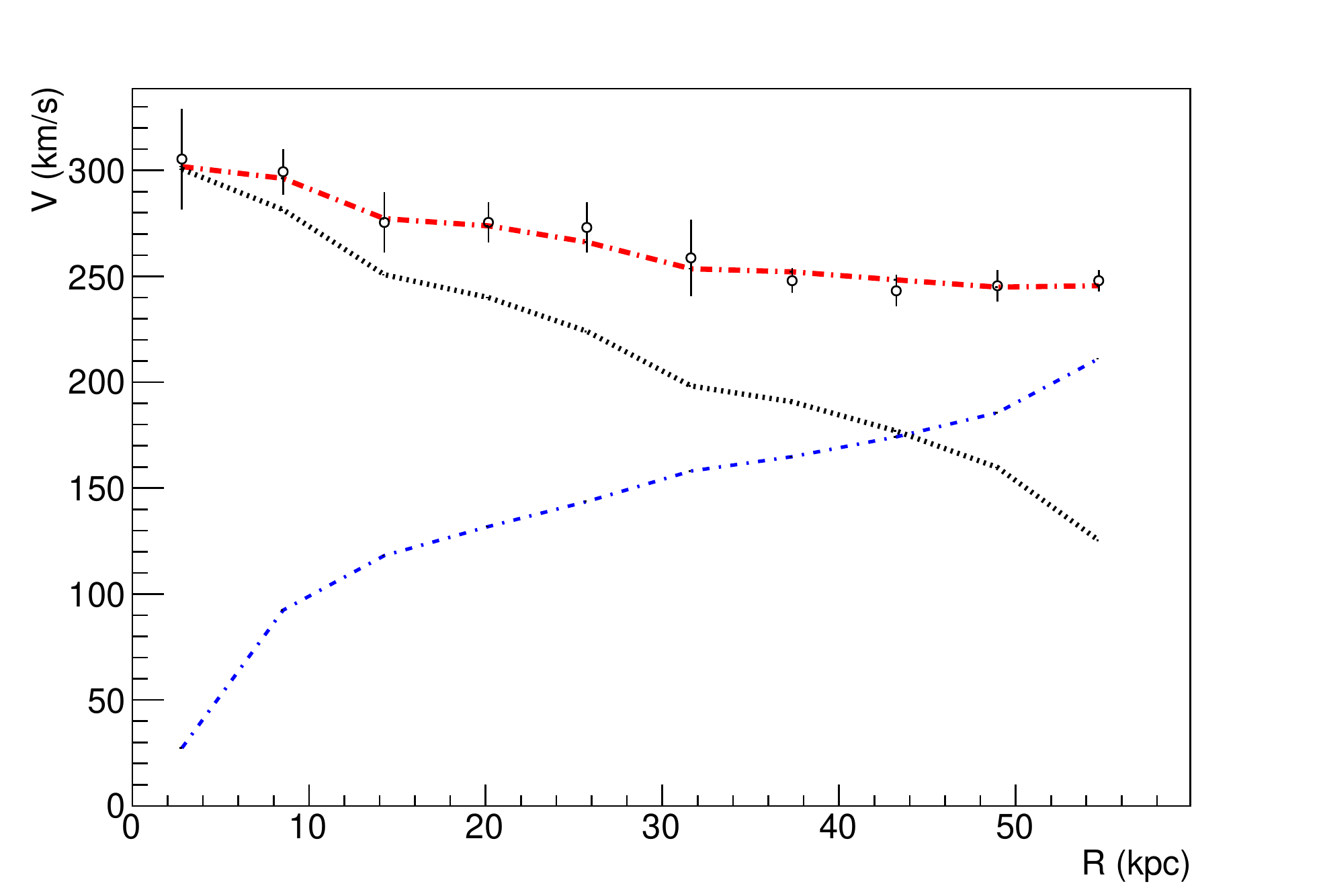}}
\subfloat[][ NGC 7814,  Ref.~11 ]{\includegraphics[width=0.33\textwidth]{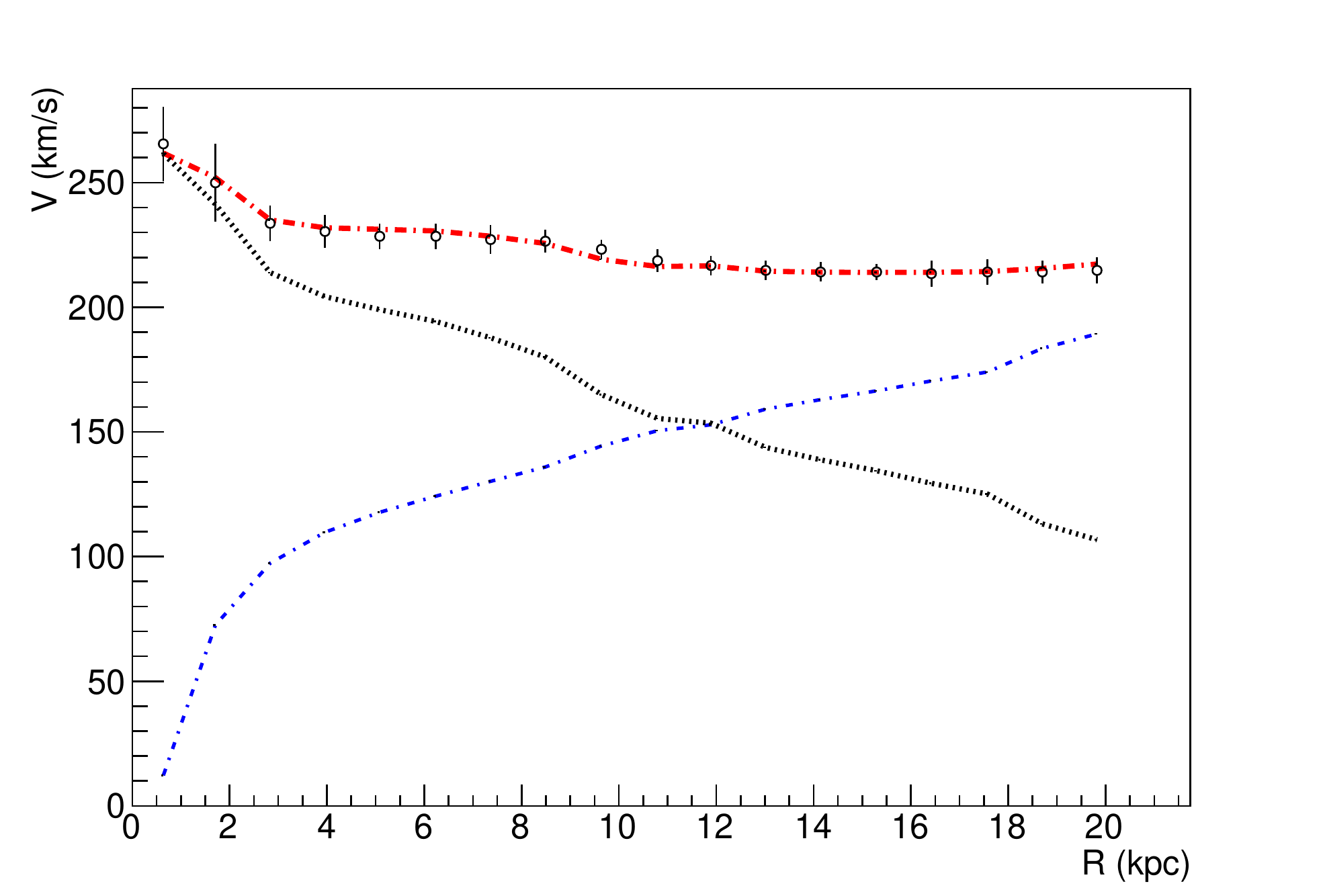}}
\\
\vspace{0.5cm} 
\subfloat[][ NGC 891, Ref.~11  ]{\includegraphics[width=0.33\textwidth]{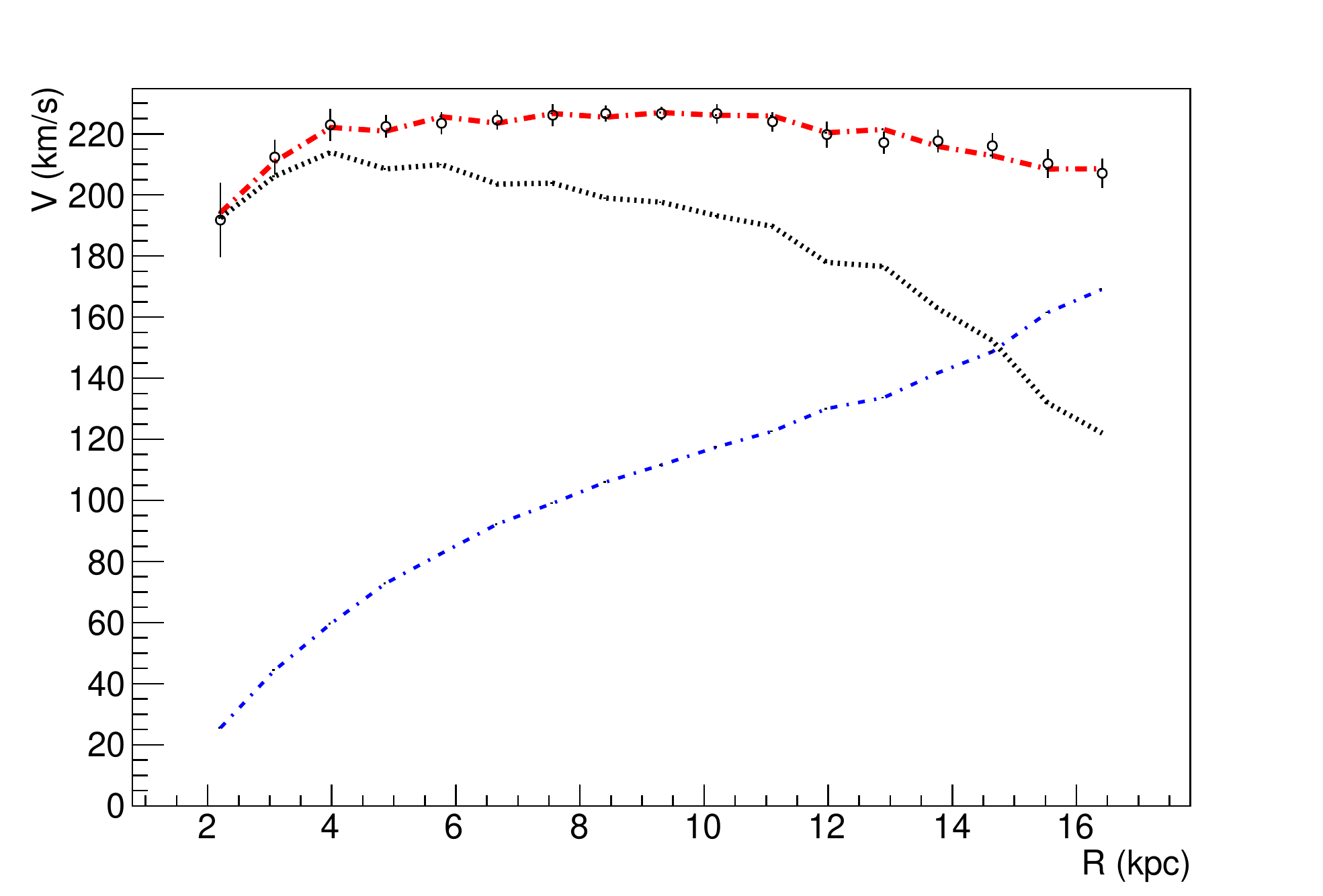}}
\subfloat[][ NGC 2841, Ref.~8 ]{\includegraphics[width=0.33\textwidth]{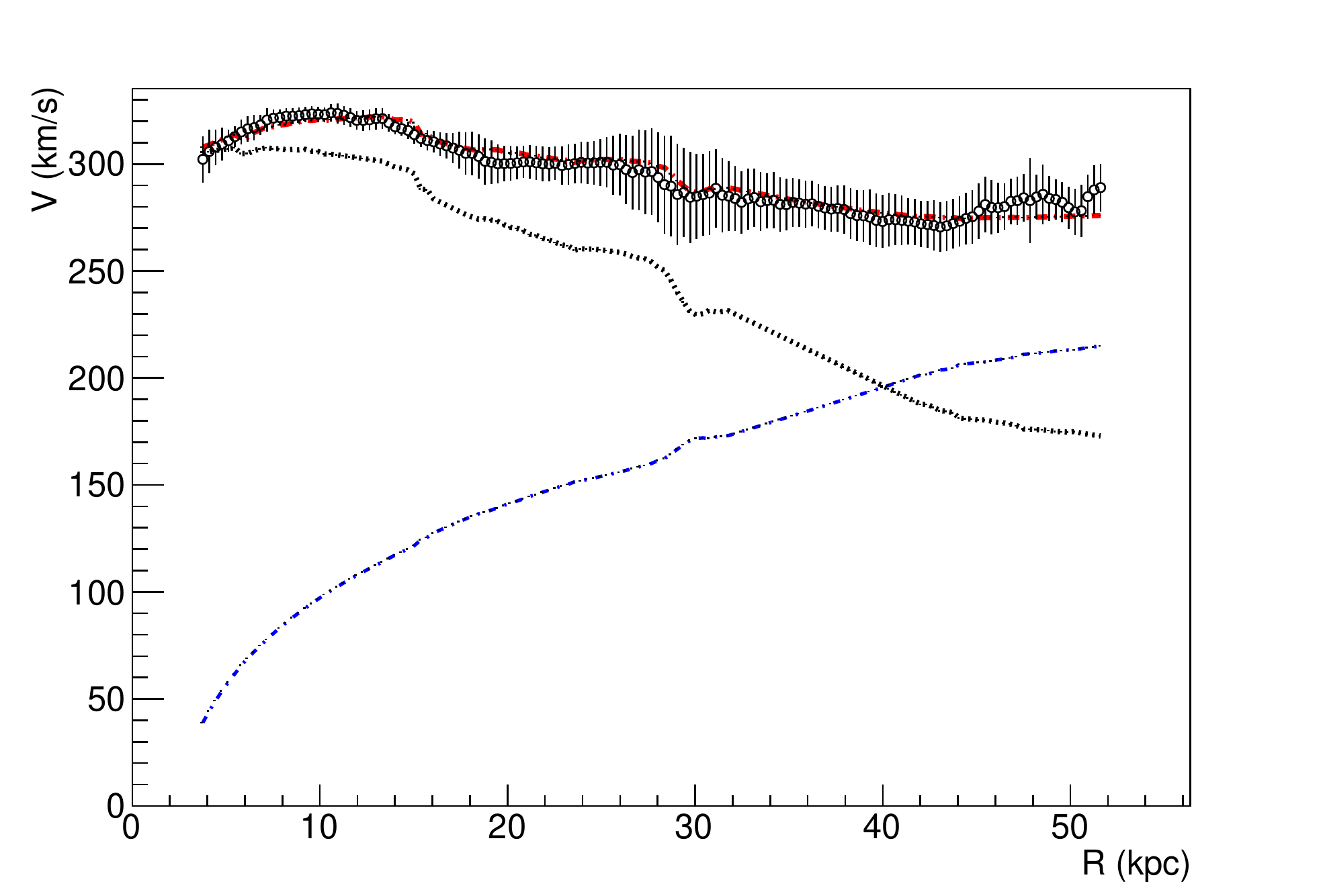}} 
\subfloat[][NGC 7331,~Ref.~8 ]{\includegraphics[width=0.33\textwidth]{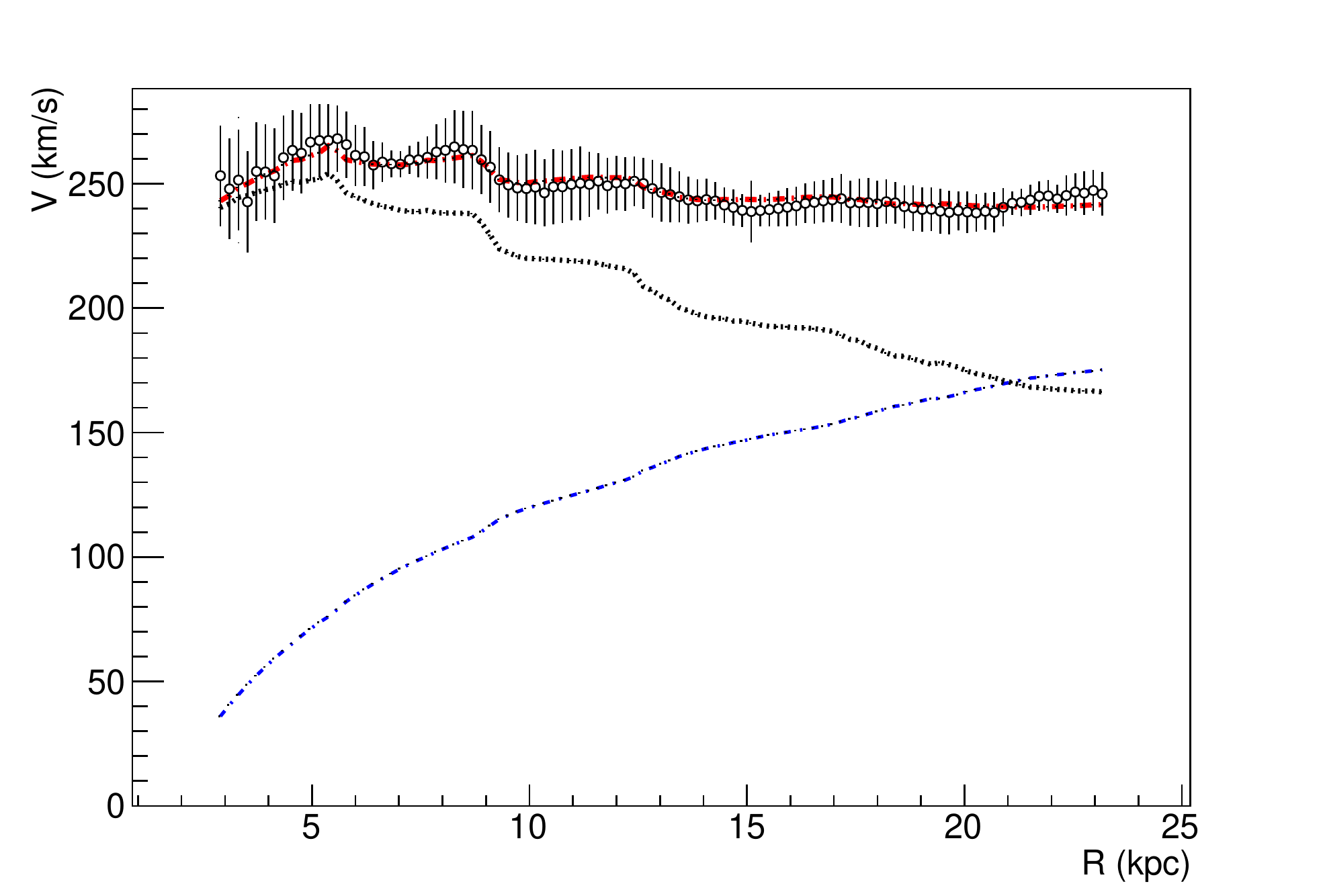}} 
\\
\vspace{0.5cm} 
\subfloat[][NGC 3521, Ref.~8  ]{\includegraphics[width=0.33\textwidth]{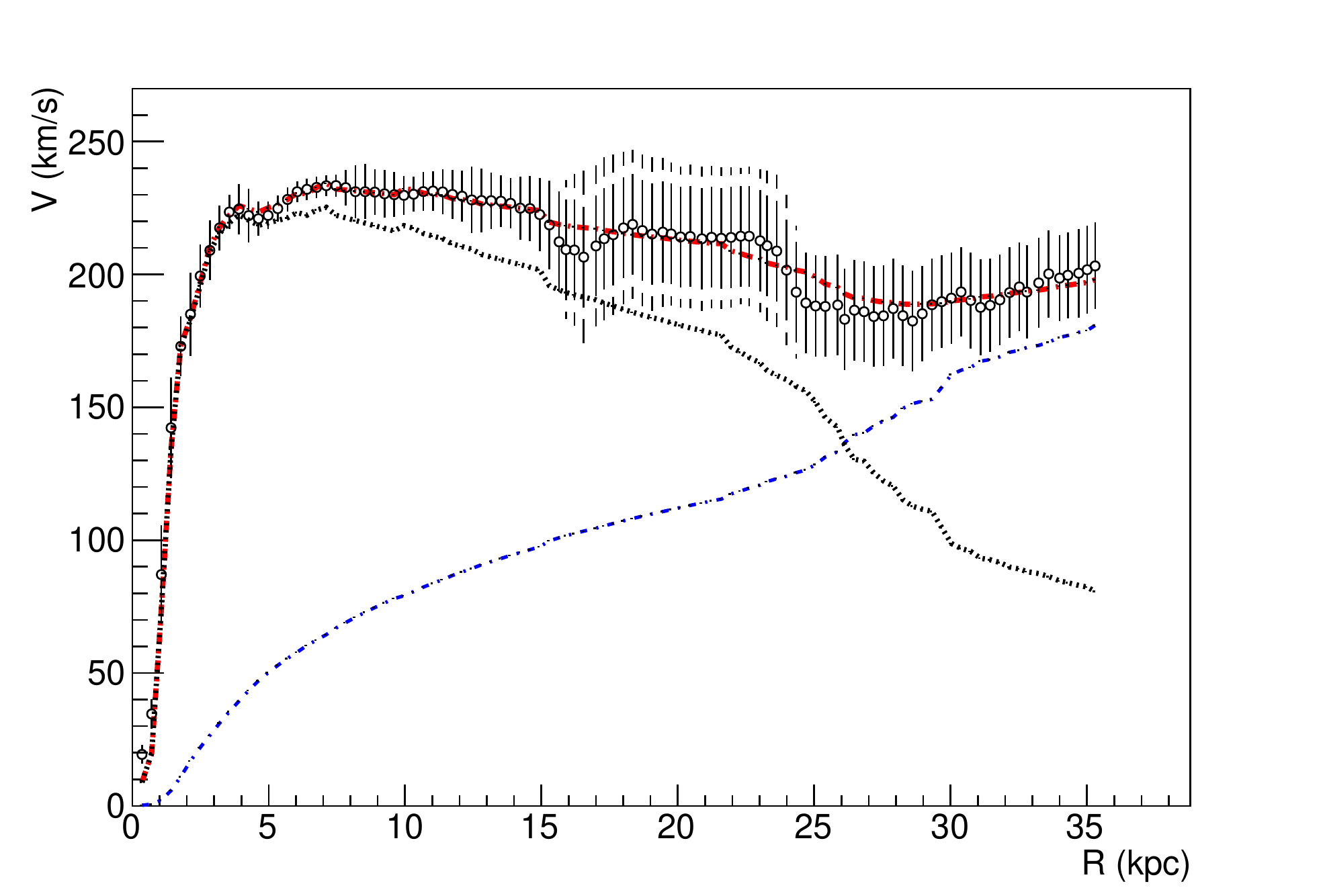}}
\subfloat[][ NGC 5055, Ref.~3   ]{\includegraphics[width=0.33\textwidth]{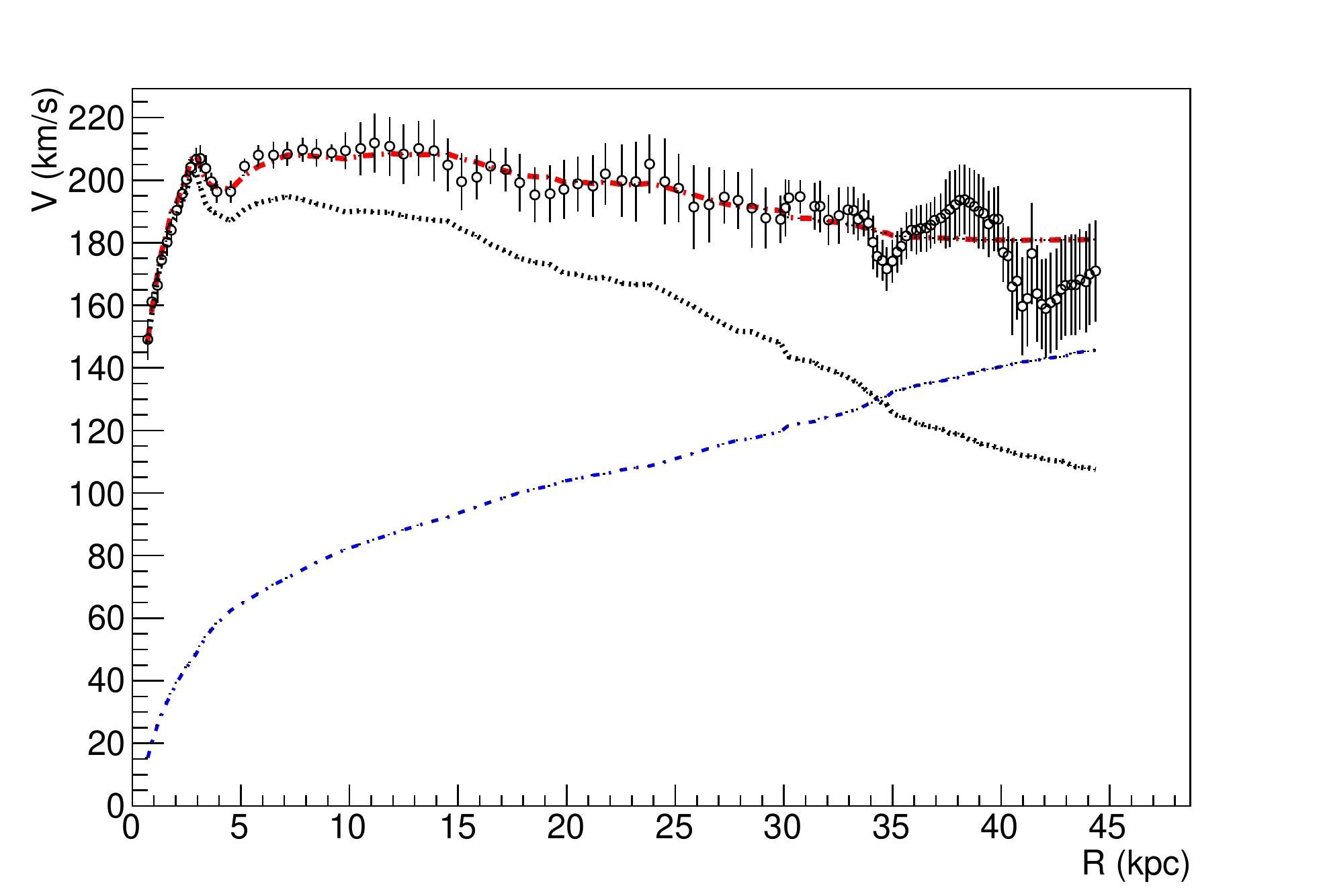}} 
\subfloat[][ NGC 4138, Ref.~10]{\includegraphics[width=0.33\textwidth]{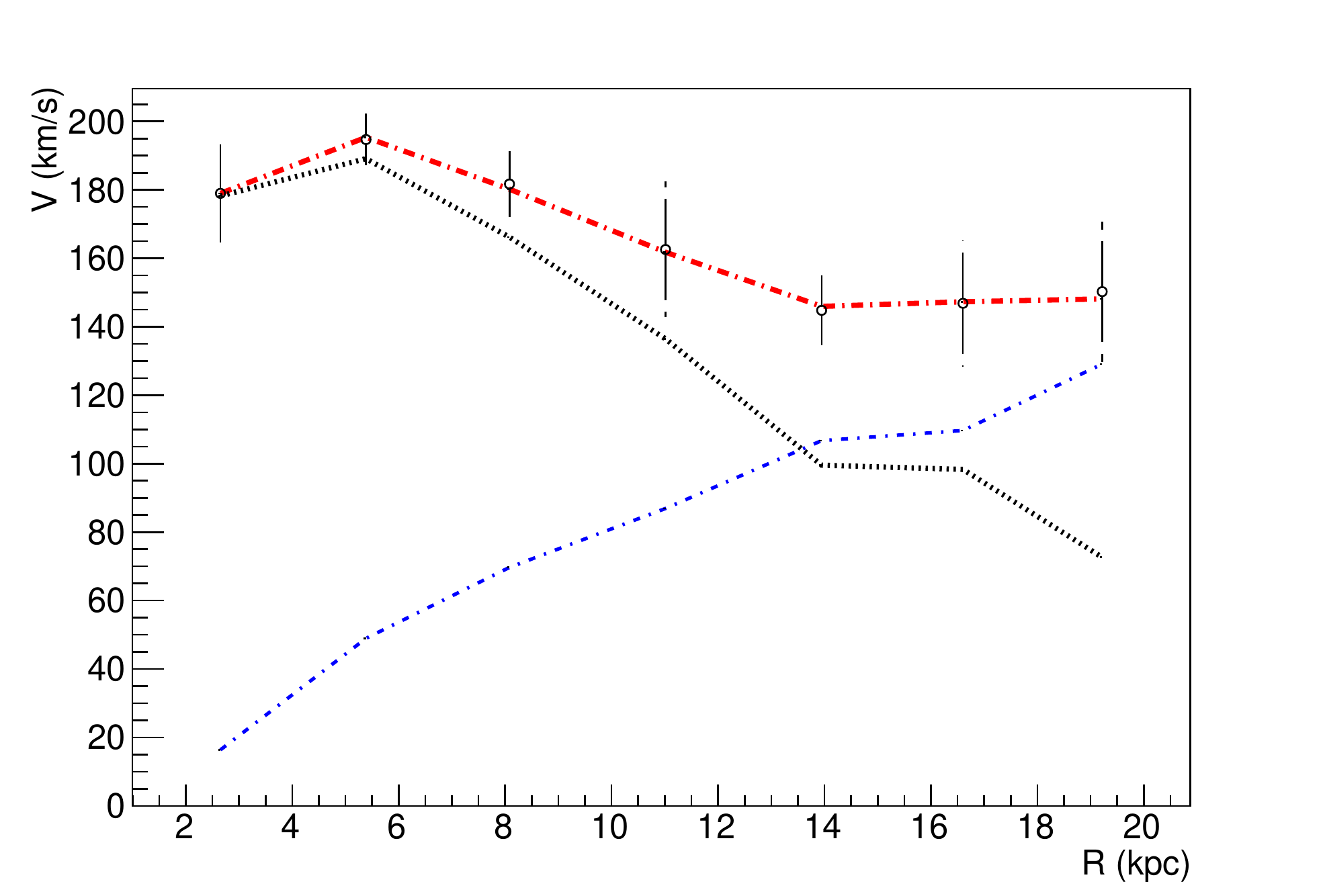}}\\
\vspace{0.5cm} 
\subfloat[][ NGC 5907,  Ref.~10 ]{\includegraphics[width=0.33\textwidth]{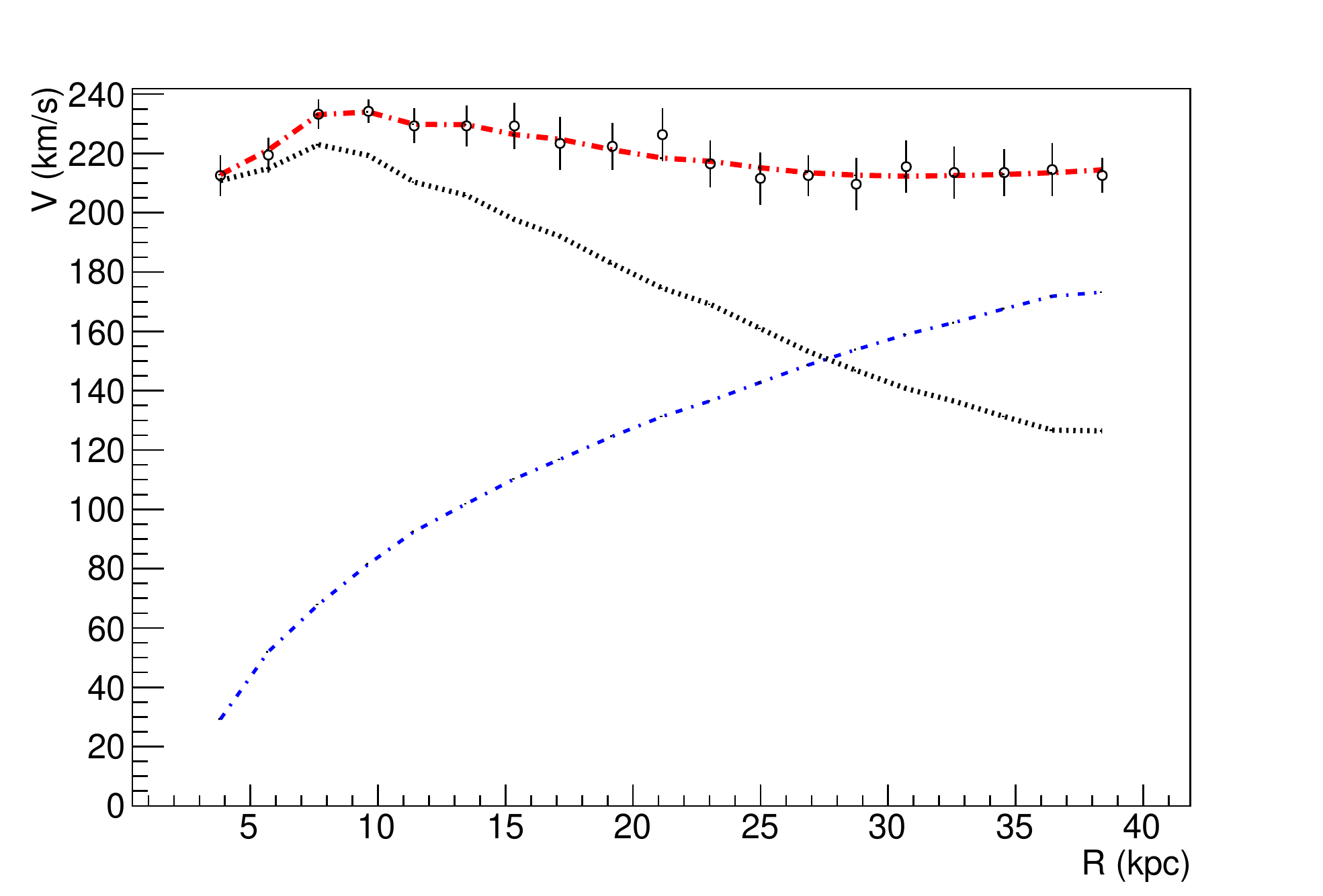}}  
\subfloat[][NGC 3992, Ref.~10 ]{\includegraphics[width=0.33\textwidth]{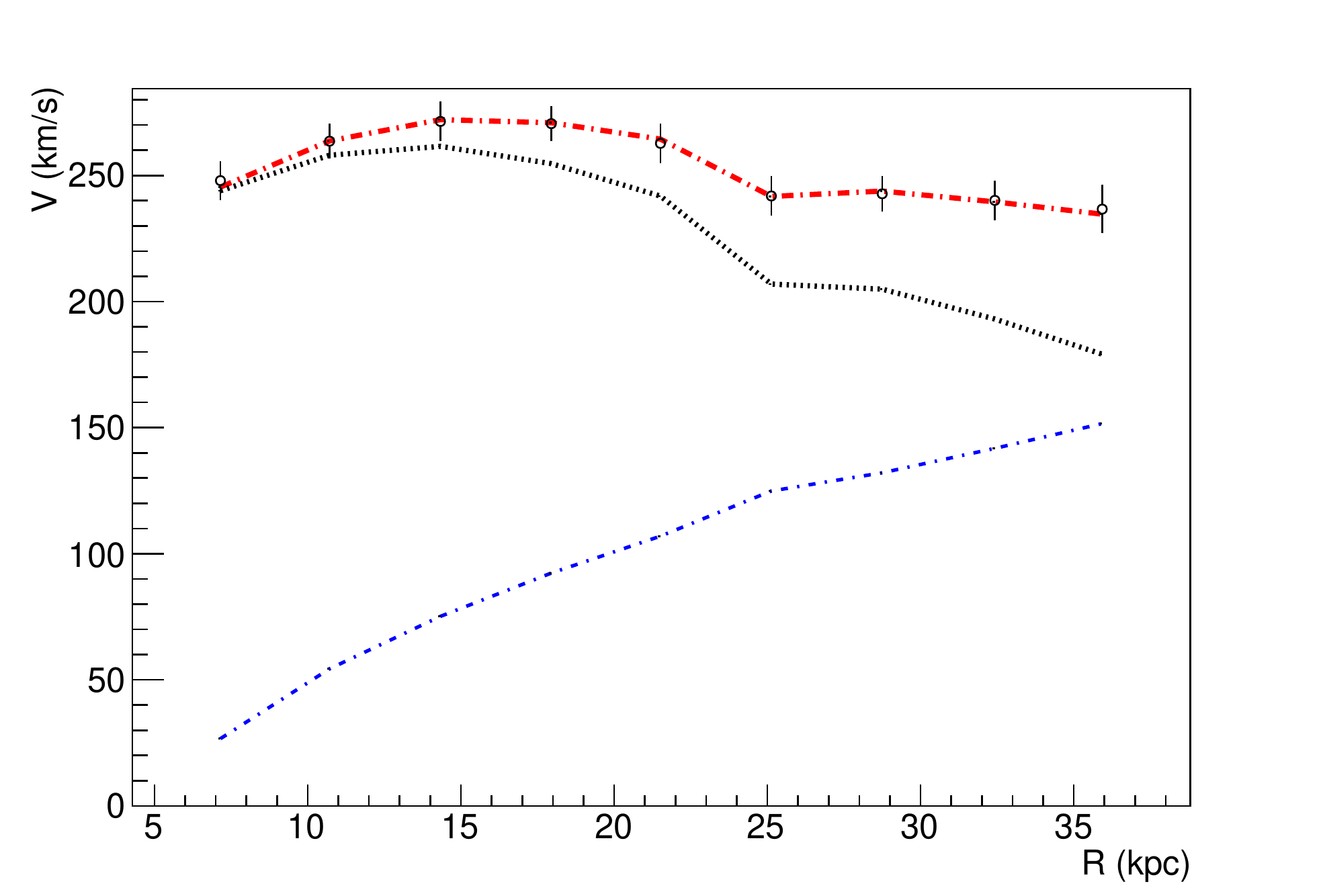}}
\subfloat[][NGC 2903, Ref.~10 ]{\includegraphics[width=0.33\textwidth]{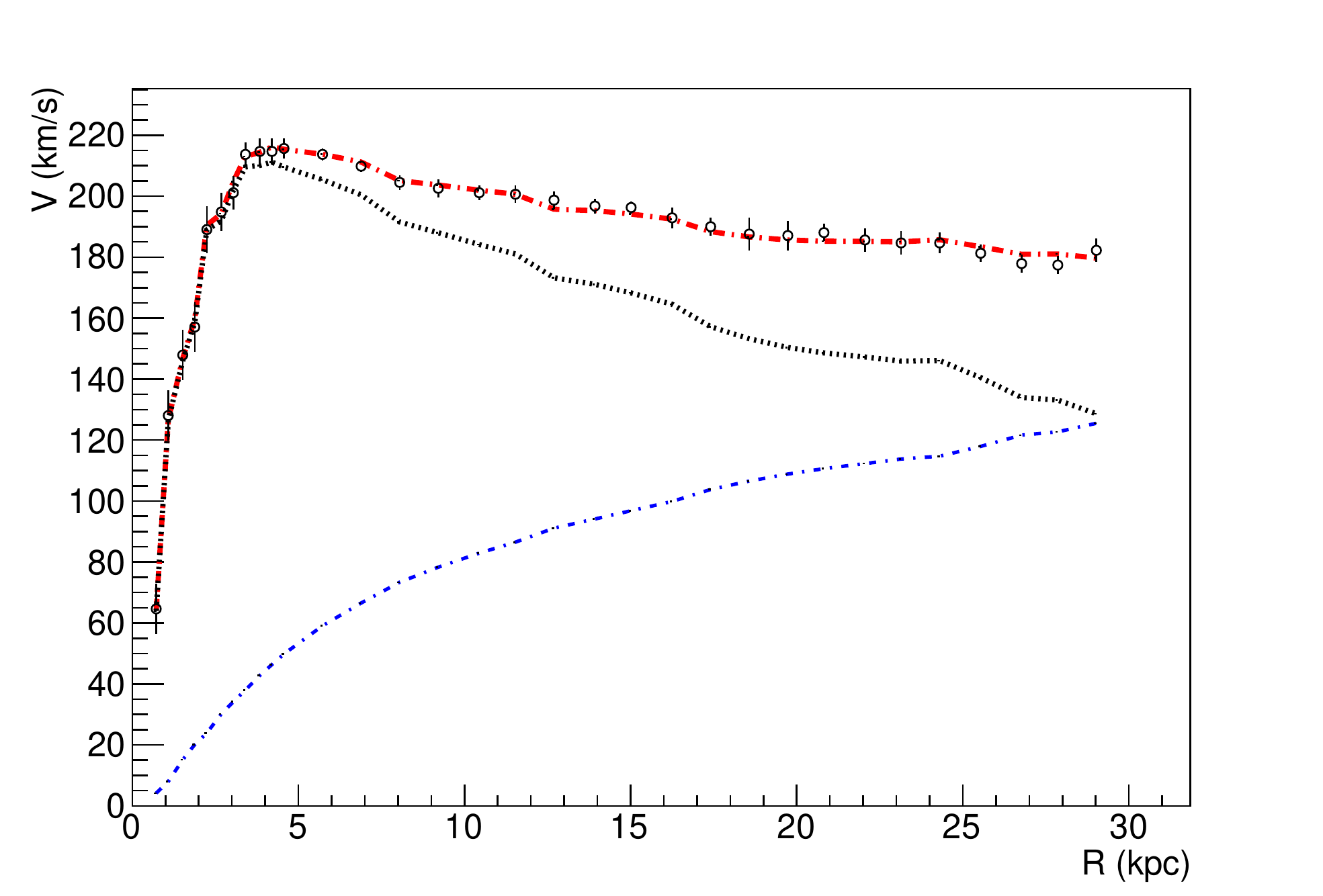}}
 \caption{LCM rotation curve fits.   In all panels  lines are;   circles with associated error bars is reported data,  red dotted-dashed is LCM fit to the reported data, blue dotted-dashed is relative curvature contribution and  the  black dotted line is the Keplerian prediction from the luminous mass.    References  are as in Table~\ref{sumRESULTS}.}    
            \label{fig:results3} 
\end{figure*}

\begin{figure*} 
 \centering
\subfloat[][NGC 6946, Ref.~10]{\includegraphics[width=0.33\textwidth]{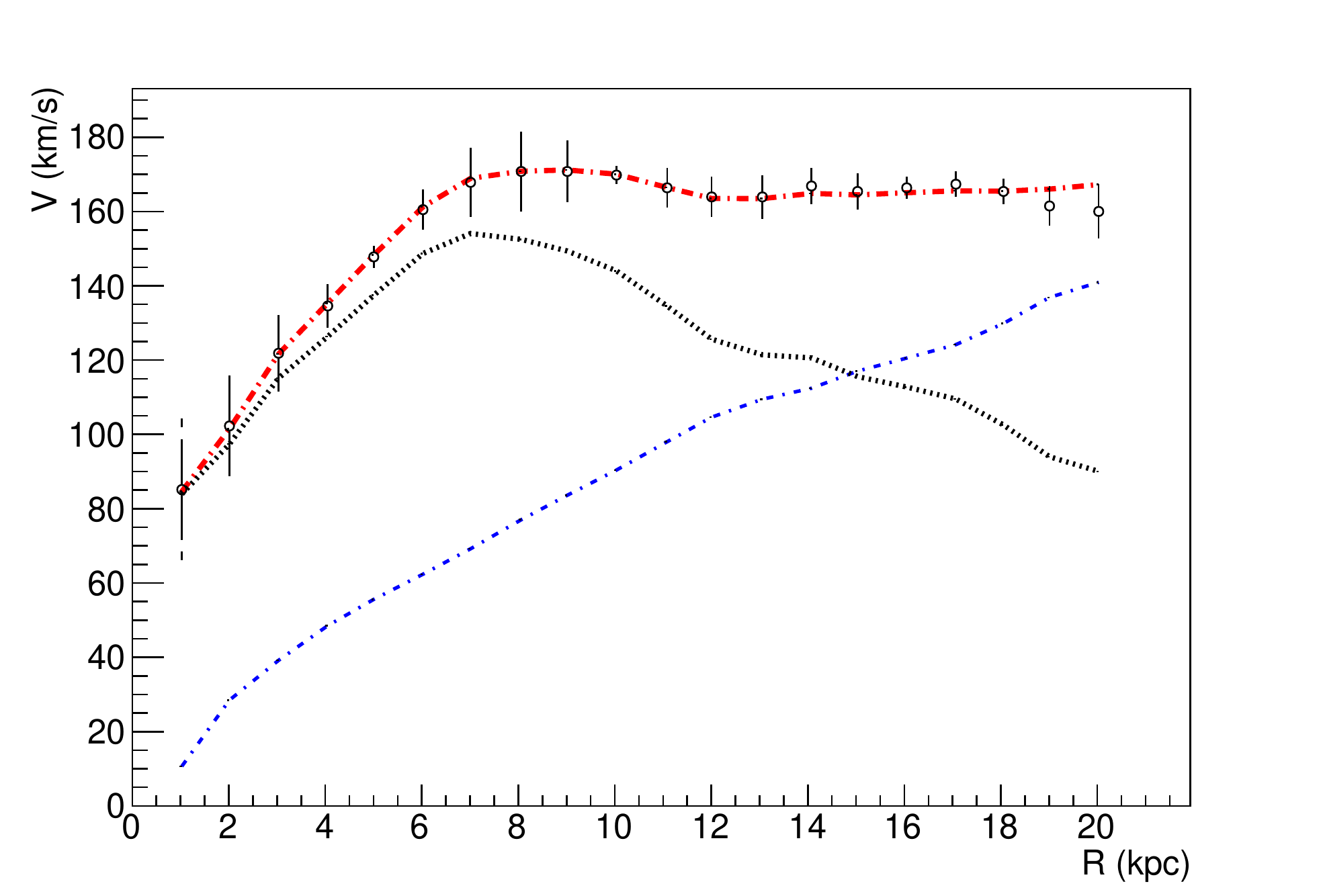}}
\subfloat[][ NGC 3953, Ref.~10 ]{\includegraphics[width=0.33\textwidth]{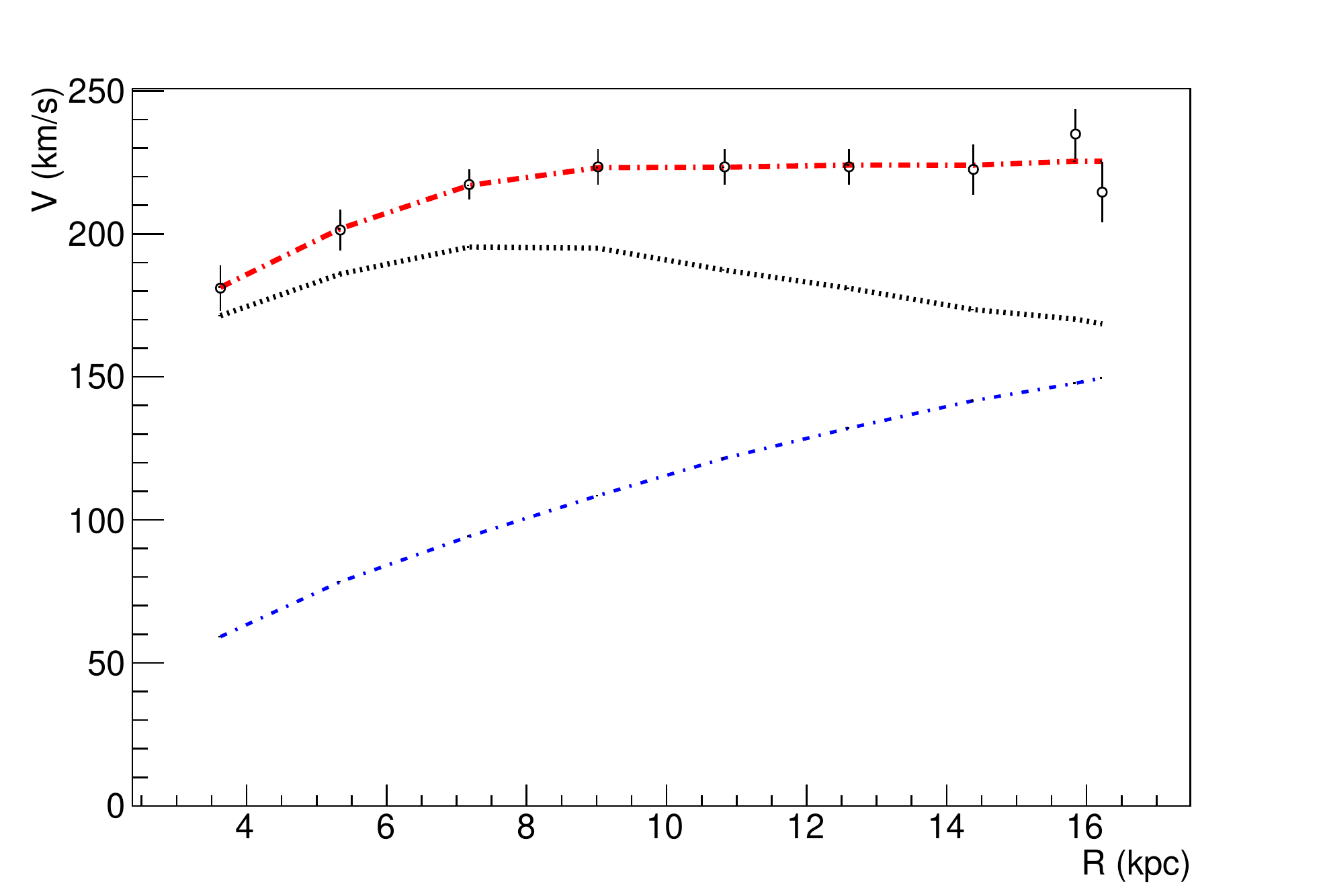}}
\subfloat[][ UGC 6973, Ref.~10 ]{\includegraphics[width=0.33\textwidth]{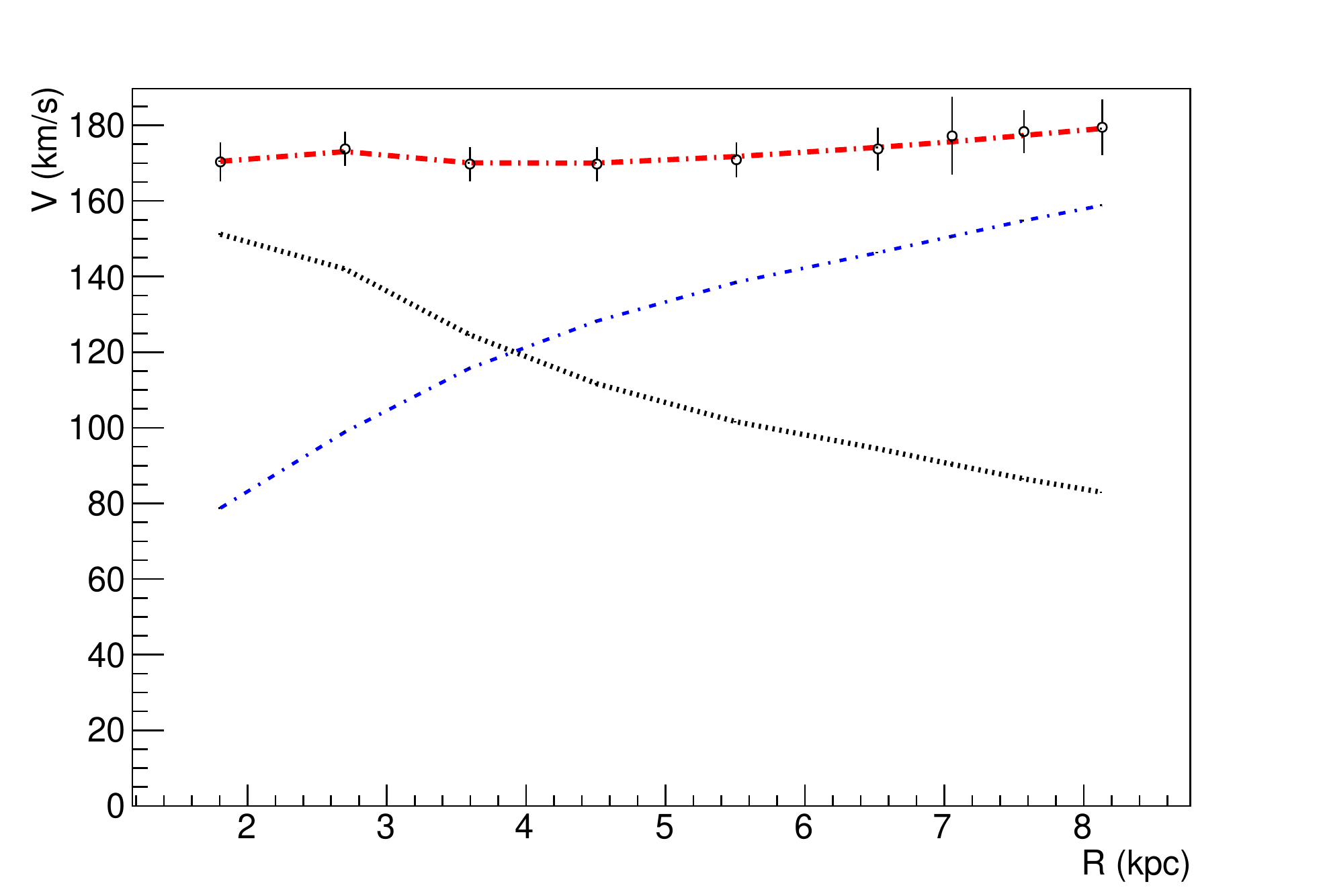}}
 \\
\vspace{0.5cm} 
\subfloat[][ NGC 4088, Ref.~10]{\includegraphics[width=0.33\textwidth]{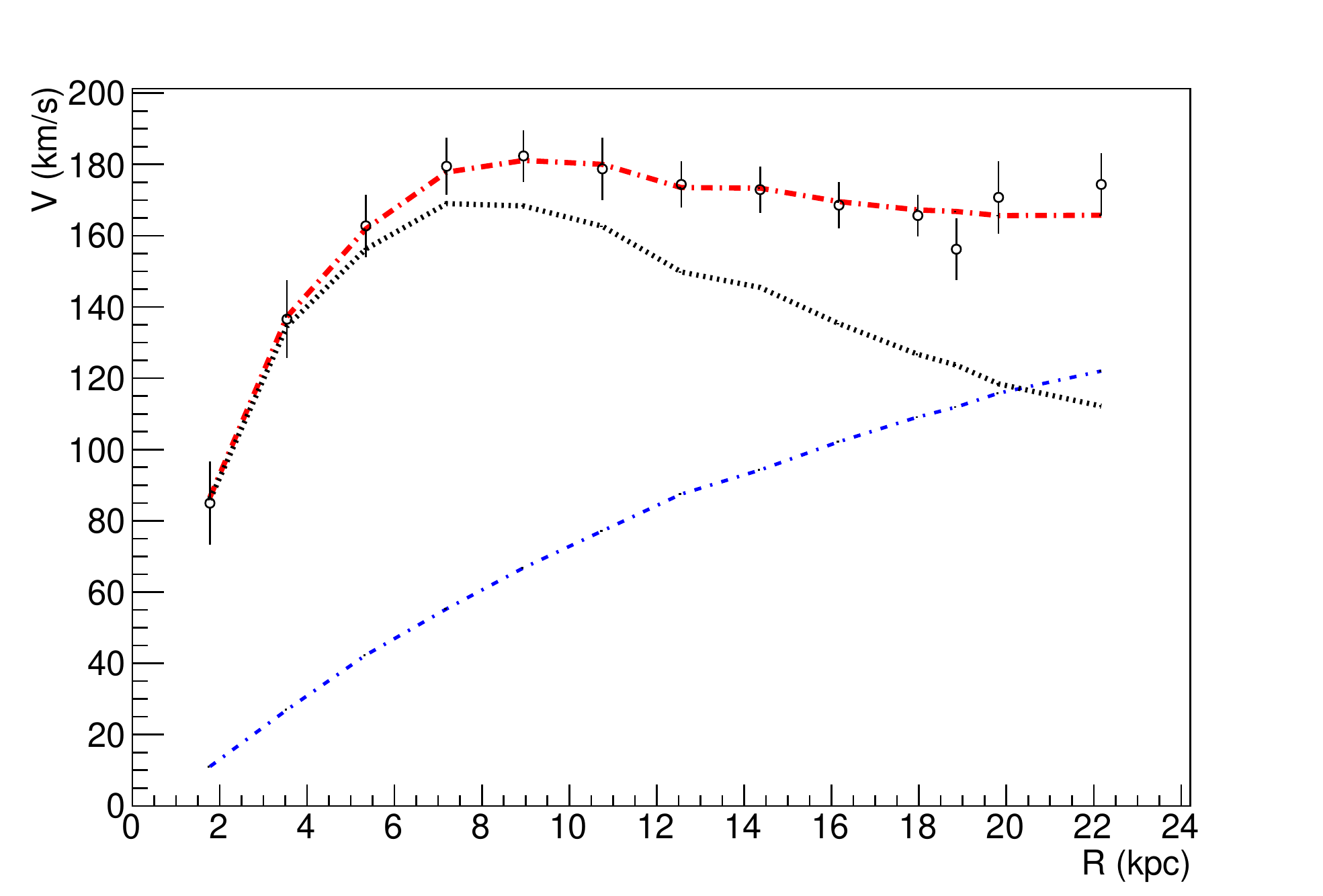}} 
\subfloat[][ NGC 3726, Ref.~10]{\includegraphics[width=0.33\textwidth]{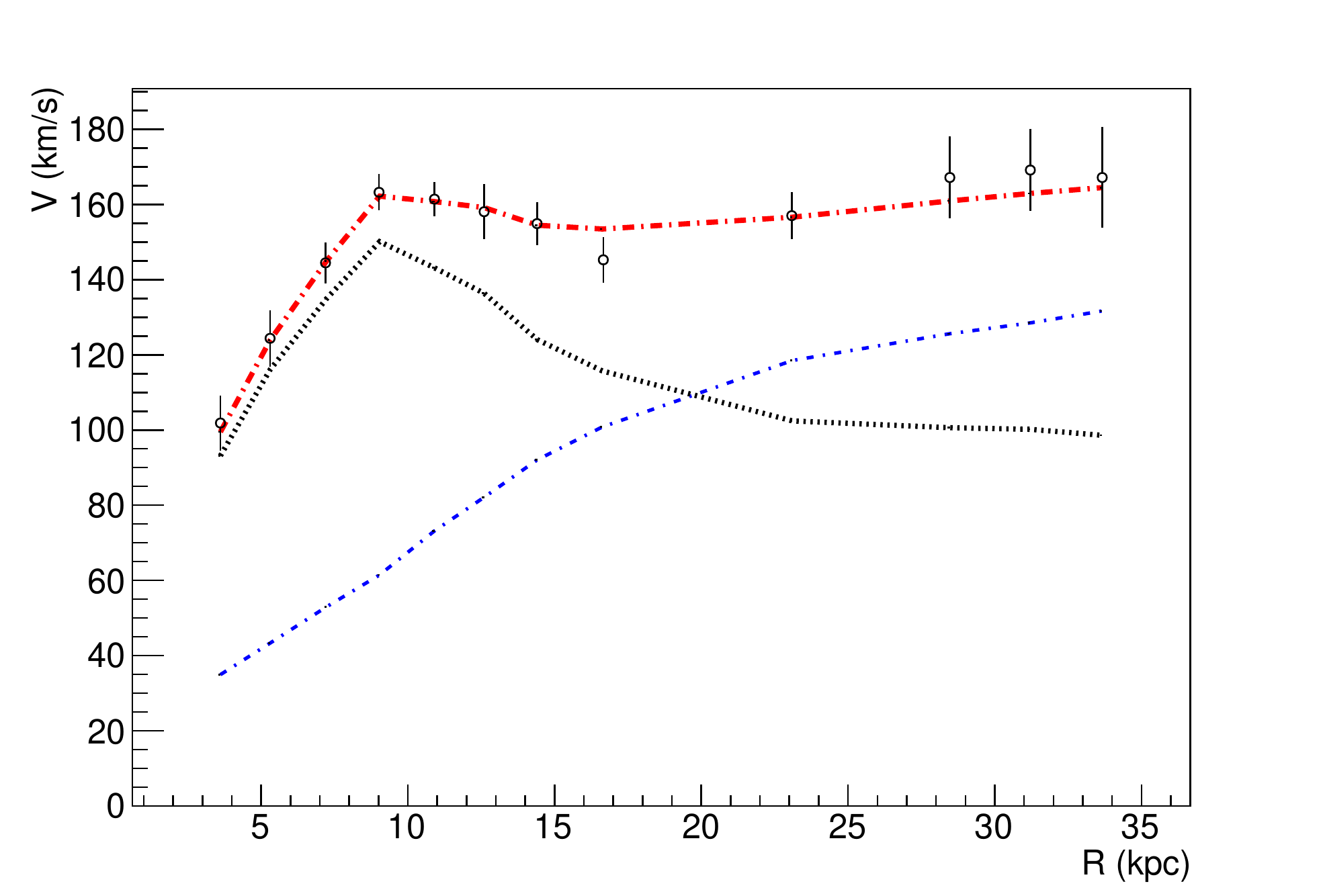}} 
\subfloat[][ NGC 2403, Ref.~3]{\includegraphics[width=0.33\textwidth]{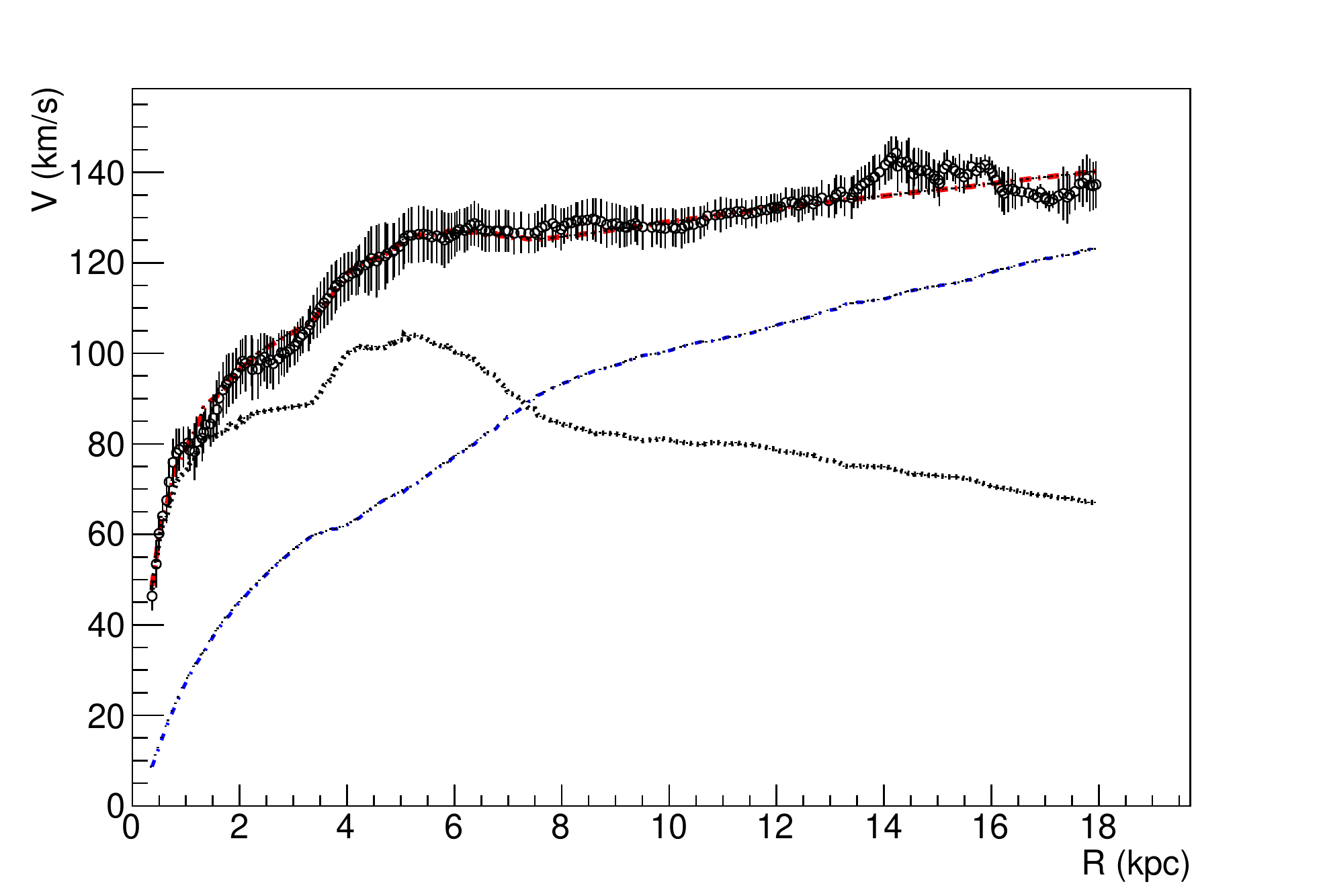}}
 \\
\vspace{0.5cm} 
\subfloat[][ NGC 3198, Ref.~3]{\includegraphics[width=0.33\textwidth]{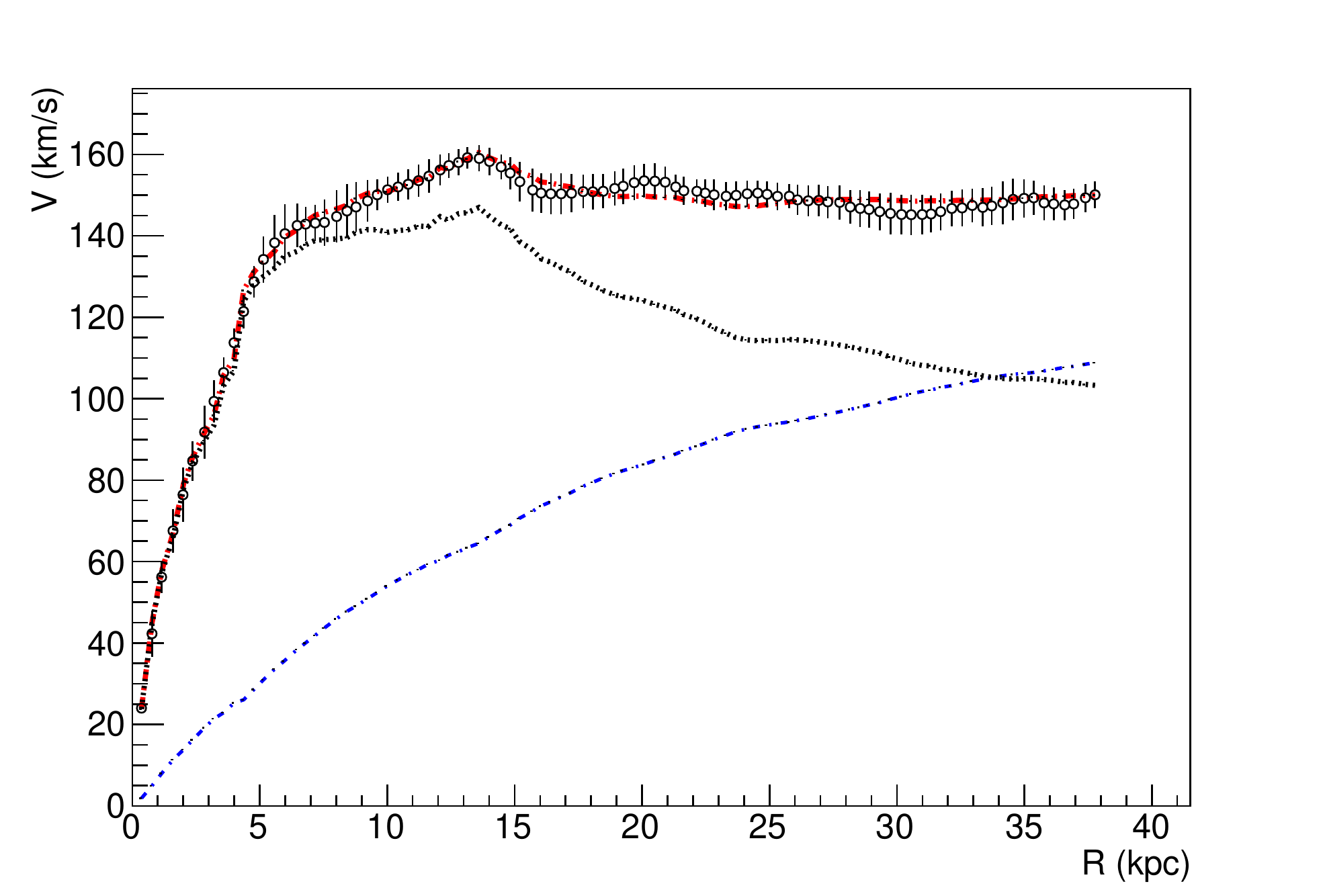}}  
\subfloat[][M 33, Ref.~5]{\includegraphics[width=0.33\textwidth]{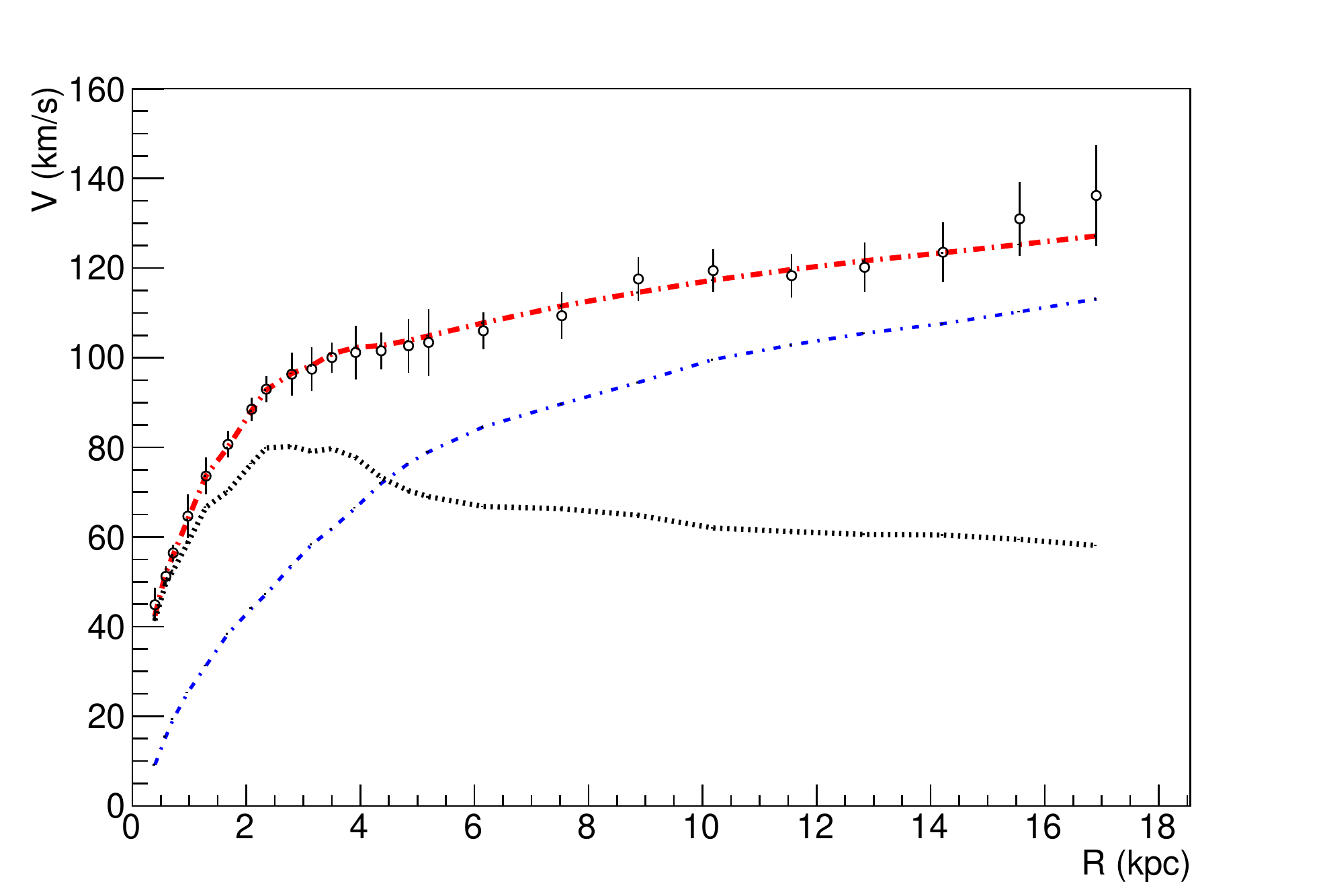}}    
\subfloat[][ F 563-1, Ref.~2]{\includegraphics[width=0.33\textwidth]{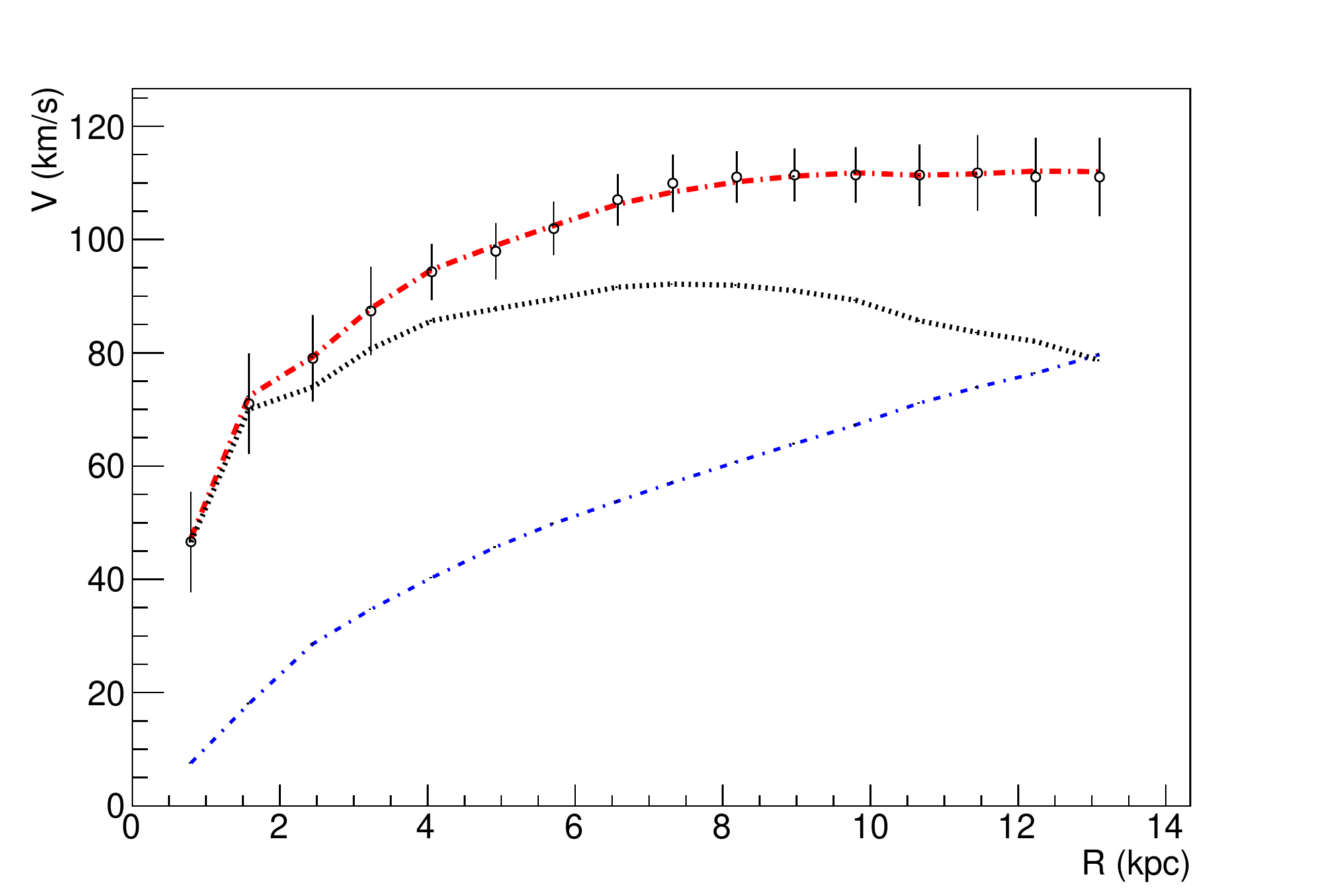}}\\
\vspace{0.5cm} 
\subfloat[][ NGC 7793, Ref.~14]{\includegraphics[width=0.33\textwidth]{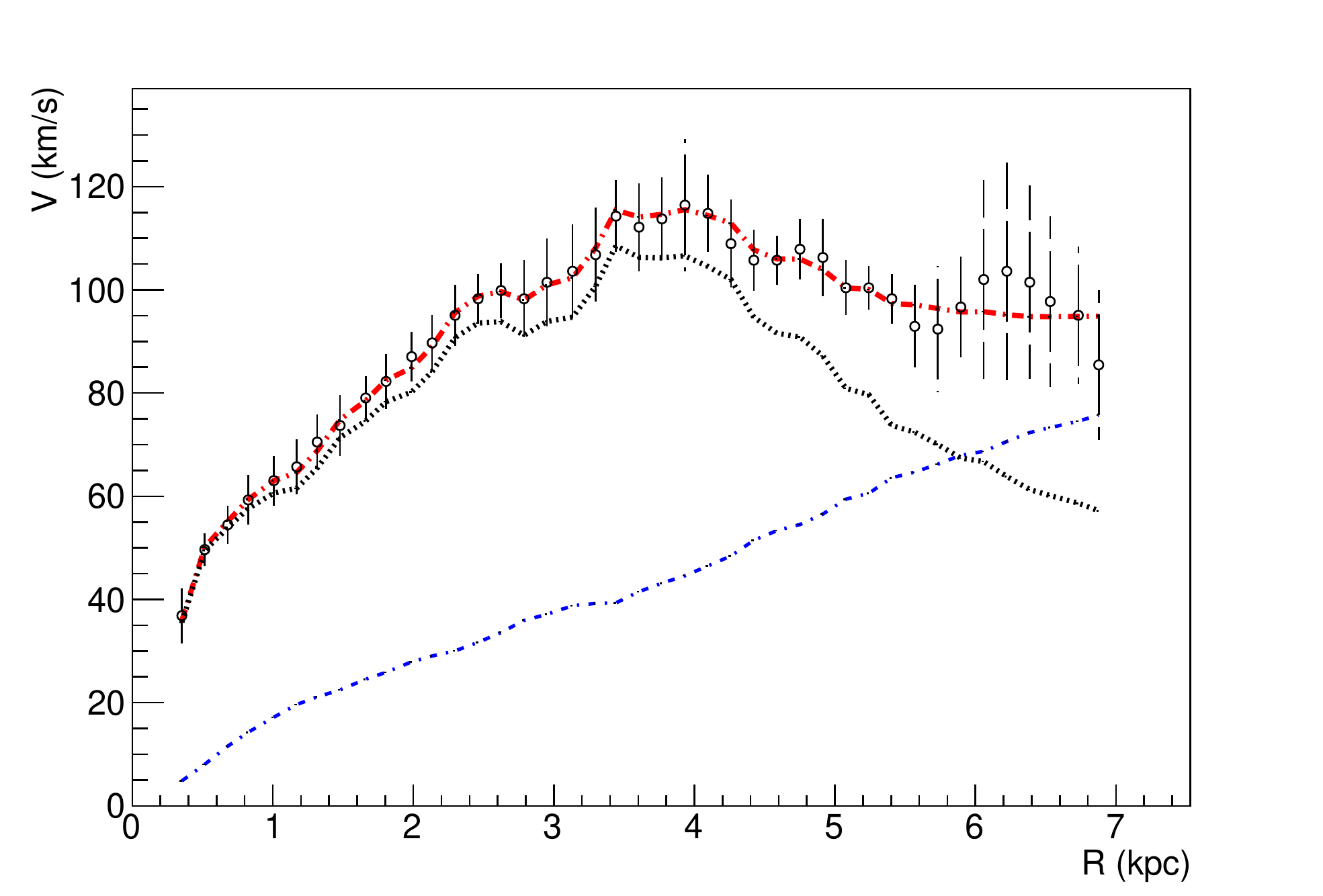}} 
\subfloat[][ NGC 925, Ref.~3  ]{\includegraphics[width=0.33\textwidth]{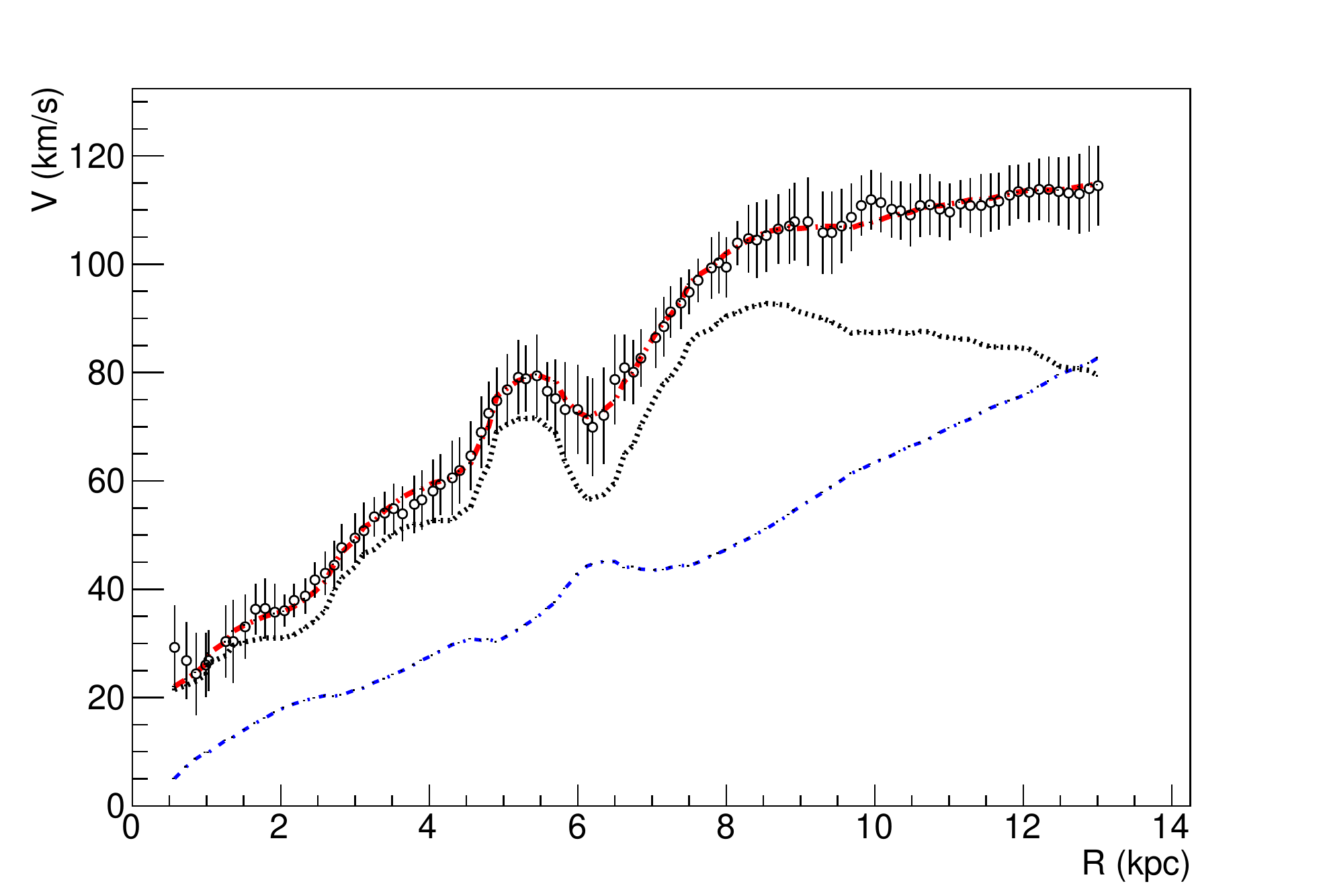}} 
\caption{LCM rotation curve fits.   In all panels:  lines are as in previous figure.   References  are as in Table~\ref{sumRESULTS}.}  
             \label{galaxiesSmallest}   
\end{figure*}

\begin{figure*} 
 \centering
\subfloat[][UGC 128, Ref.~6]{\includegraphics[width=0.33\textwidth]{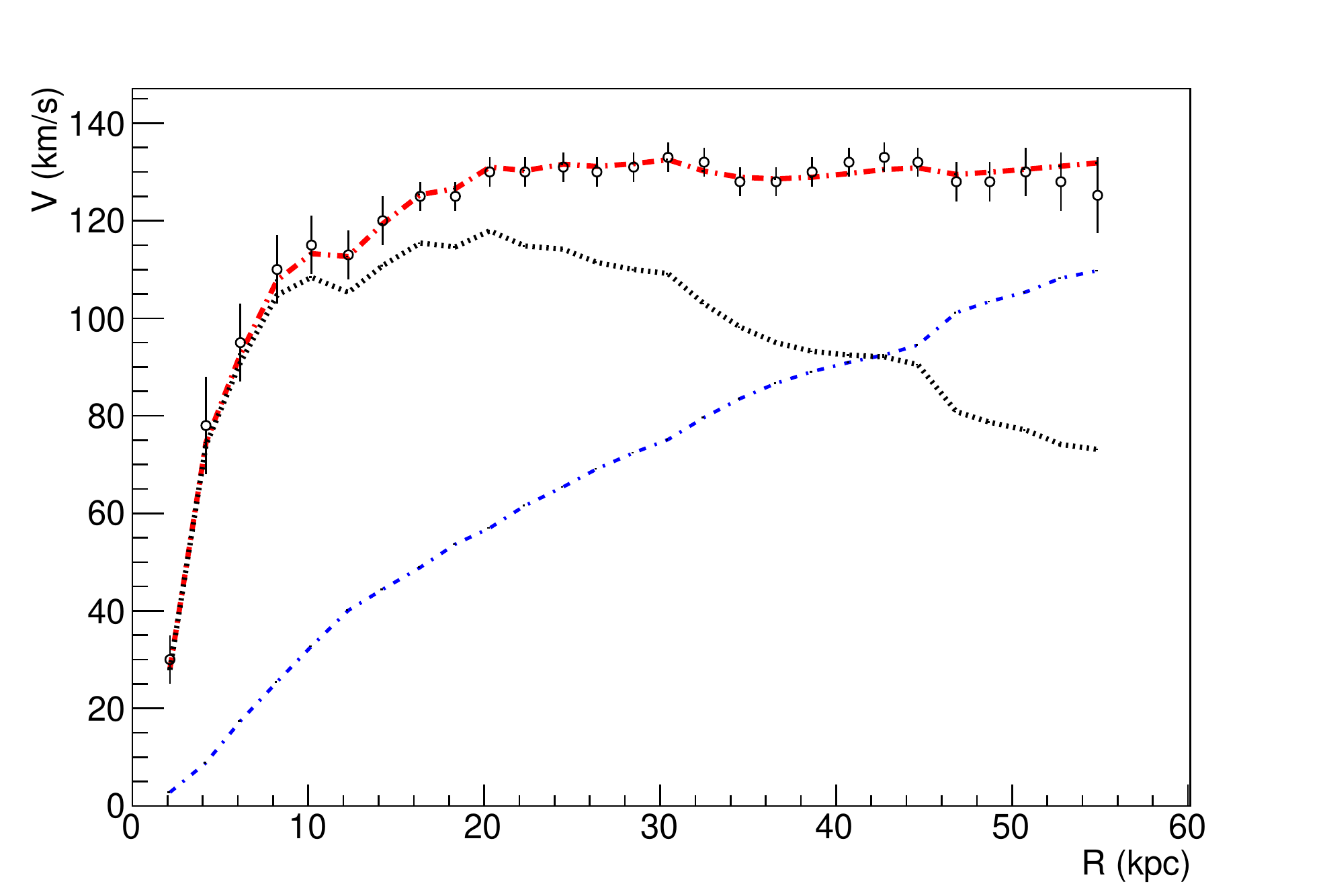}}
\subfloat[][UGC  7524 , Ref.~6]{\includegraphics[width=0.33\textwidth]{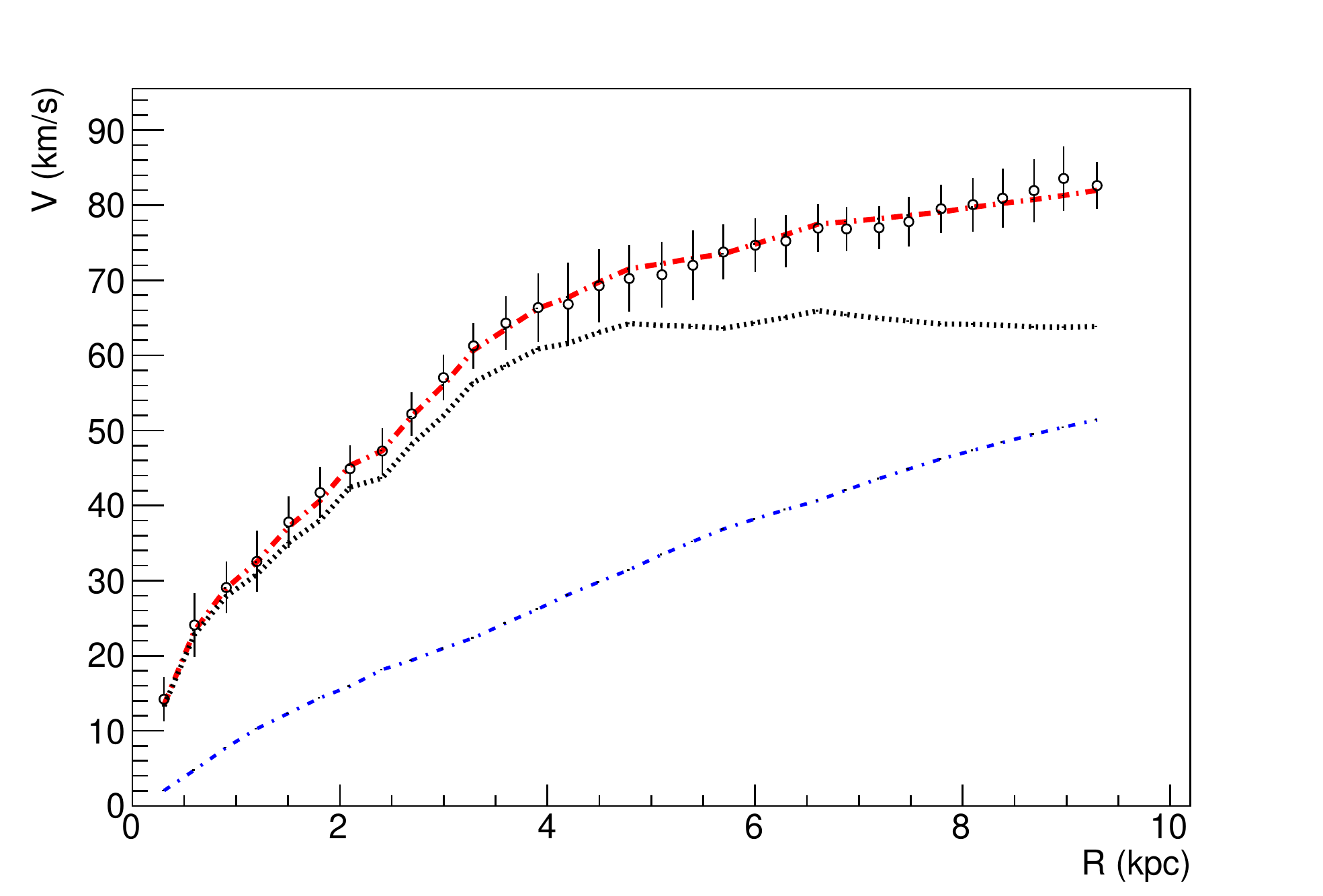}}  \\
\caption{LCM rotation curve fits.   In all panels:  lines are as in previous figure.   References  are as in Table~\ref{sumRESULTS}.}  
             \label{galaxiesS}   
\end{figure*}

   The identification of the LCM free parameter $\alpha$ as highly correlated with the function Eq.~\ref{correl} can now be explored.   We fit the distribution of  the LCM $\alpha$ results against the  emitter galaxy radial densities  (defined in Eq.~\ref{radDens}) with a power law.  The power law fits result  in an  average value for the exponent of $1.64$ (see Figure~\ref{fig:alphaDistrib1}). While it is notably coincidental that this  exponent is   similar to those often   reported for  the  dark halo density parameters in   NFW models~\citep{NFW}, we do not in this paper make any comments or claims to any relationship with the NFW.  However, since all DM or alternative gravitational theories are seeking to explain the same physics, via differently described phenomena, coincidences as such are both unavoidable and beneficial to the entire community to spark dialogue of overlap and plausibility. 
 The result of the fits to the $\alpha$ distribution for the entire sample has given the LCM a way to predict the scale length of the given MW model, given that we use the masses reported for each model. 
 The LCM prediction for the  MW scalelengths are reported in  Table~\ref{MWlum},  along with the corresponding    $\chi^2_R$ value for each MW    power law fit.  We note that while the Xue/Sofue MW is not the best $\chi^2_R$ value, it does fit all  emitter galaxies  in the sample  without artificial constraints on the bounds for the $\alpha$ parameter. 

 \begin{table}  \centering
  \begin{minipage}{140mm}
  \caption{ Milky Way Luminous Mass Models \& $\alpha$ analysis  
\label{tab:MWlum} }
  \begin{tabular}{@{}llccc c l r@{}}
  \hline
 Galaxy     	 & $R_{last} $ &$M_{bulge} $&$M_{disk} $&$ b$ & $r_e$ & $\chi^2_R$ & $ dof $\\ 
 \hline
 Xue/Sofue &60kpc	&$1.8$	&$5.30$ &$1.611$     &3.74&13.19& 24\\
 Sofue 	 & 30 kpc 	& $1.8$	&$6.80$  & $1.683$    &4.76 &6.52&23\\
 Xue 		&60 kpc	&$1.5$	&$ 5.00$ &   $1.523$ &5.76&11.45&24\\
 Klypin,  A	&15 kpc	&$0.8$	&$4.00$ &  $1.699$    &3.87&1.76&20\\
 Klypin,  B  	&15 kpc	&$1.0$	&$5.00$ &   $1.696$   &4.53&2.65&20\\
\hline 
\end{tabular}\\ 
 Masses in   units  of $10^{10}M_{\odot}$,
 exponential scalelengths $r_e$ in ($kpc$), predicted \\
from  fit values of $b$ , for   power law fits of the form  
  $\alpha =  \left(\rho_{mw}/\rho_{gal}\right)^b $.  \\
  Goodness of fit indicated by reduced $\chi^2_R$ values per  degrees of freedom ($ dof $).
 \\
 \end{minipage}
  \end{table} 
 \begin{table*}
 \centering
 \begin{minipage}{140mm}
  \caption{Results   for originating   model and   LCM  
  \label{sumRESULTS}}
  \begin{tabular}{@{}lrrrlrrrrrc@{}}
  \hline
   Galaxy     	  &Ref.~& Distance& Luminosity& \multicolumn{2}{l}{Other fit results}	& & \multicolumn{4}{l}{\underline{LCM fit results}}  \\
\hline
   	 	& &(Mpc) &$10^{10}L_{\sun}$ & 	 Model & $\chi^2_r$	&&$\alpha$&$\rmn{M/L}$&$r_e$&$\chi^2_r$ \\ 
 \hline
F 563-1	& 2	&45&0.15
					&NFW	&0.05 & &5.53	&12.53&2.60 &0.02 \\
M 31* 	& 12 &0.78&2.60	
					&ISO		&0.36 & 		&0.65	&5.96&4.80 &0.04  \\
M 33		& 5	&0.84&0.57 
					& NFW 	&2.46&& 13.01 &0.88	 &1.46	& 0.14 \\
NGC 891*	& 11 &9.5&2.50	
					&MaxLight  &1.10 & &0.94	 &3.11  	 &4.14	&0.25 \\ 
NGC 925 	&3	&9.25&1.61
 					&ISO 	&2.40& 	&23.10&0.85	 &4.35	&0.07\\
NGC 2403	 & 3	&3.22&0.93
					& NFW	&4.56&& 6.25	&1.42&2.18& 0.56 \\
NGC 2841*  &6	&14.1&4.74
					&CG		&2.27 &	&0.21&5.46 &3.77	&0.11 \\
NGC 2903  &10 &6.4&3.66	 
					&MOND	&10.71& &0.39&2.03&2.53&0.31\\
 NGC 3198 & 3 &13.8&3.24	
					 &NFW	&5.40  &   &1.55	 &1.88	 &4.41	&0.26   \\
NGC 3521  & 6&10.7&4.77	 
					&CG		&1.37& &0.62	 &2.15  &3.29&0.22 \\
NGC 3726	& 10&18.6&3.77	
					&MOND	 &3.57&& 2.86	&1.04&4.02	&0.27 \\
NGC 3953	& 10&18.6&4.19	
					&MOND	 &1.35&	 &0.97	&1.93&3.36 	&0.32 \\
NGC 3992	& 10&18.6&7.50	
					&MOND	 &0.50& &0.51&2.34  &4.66&0.03 \\
NGC 4088	& 10&18.6&0.88	
					&MOND	 &1.70& &1.52	&6.41&3.65	&0.27 \\
NGC 4138	& 10&18.6&1.18	
					&MOND	 &2.12&	 &	1.19&2.93 &1.55	&0.01 \\
NGC 5055*	 & 3	&10.1&3.62
					&  NFW	 &17.23&  &0.31	&3.01&3.30	&0.53 \\
NGC 5533*	 & 10&54& 4.5
					&MOND	 &1.57 & &0.19	&7.58 &7.04	&0.22  \\
NGC 5907*	& 10&18.6&7.20	
					&MOND	 &0.44& 	&0.73	&1.87 &5.05	&0.10 \\
NGC 6946*	&  10&6.9&2.73	 
					&  MOND	 &   3.03& &2.73	&1.42&3.04	&0.14\\ 
NGC 7331	&6&14&6.77	
					&CG		 &1.24& &0.50	&1.48&2.98	&0.10 \\
NGC 7793  &14	&3.38&0.31
					&ISO		 &1.08& 	&5.56&2.51&1.15&0.06 \\
NGC 7814* &11 &14.6&1.30	
					&ISO   	 & 0.25& 	&0.32&4.38 &1.37&0.15 \\ 
UGC 128   &6	&64.4& 4.60
					&CG		 &1.08	&		&4.50	&2.62
					&8.14	&0.19\\
UGC 6973	& 10&18.6&0.89
					&MOND	 &23.5 & 	&2.01&1.27 & 0.86&0.02 \\
UGC 7524	& 6	&4.12&0.37
					&CG		 &0.39 & 	&16.30&2.23 &3.32 &0.06\\
\hline 
\end{tabular}
\\
     Values in this table  are  reported for the  galaxy  pairings to the Milky Way synthesized from  \cite{Xue}  and \cite{Sofue}.    Mass-to-light ratios   in units of $M_{\sun} / L_{\sun}$.   Reduced $\chi^2$ per degree of freedom indicated by  $ \chi^2_r$.  Galaxies for which the original literature reported a bulge component are indicated by (*).  Other models include the dark matter halo models (NFW and ISO),  Conformal Gravity (CG), MOND  modified Newtonian Dynamics (MOND) and  maximum disk (baryonic) mass model (MaxLight).\\
{\bf References:} 
 1.~\cite{Bege}, 2.~\cite{JNav}, 3.~\cite{Blok} , 4.~\cite{Maria}, 
5.~\cite{Cor03}, 6.~\citet{James},   7.~\cite{Batt},   8.~\cite{Gent},   9.~\cite{Bot},   10.~\cite{SanMcGa},
  11.~\cite{Frat},   12.~\cite{Car},   13.~\cite{giraud2000universal},   14.~\cite{Dicaire}. \\
    \end{minipage}
\end{table*}





\section[]{Conclusions}
\label{sec:conclusion}

We   posit that what is presented in this paper will help to constrain the Milky Way mass models. The only free parameter $\alpha$ in  Eq.~\ref{correl}, has been shown to high confidence to be the ratio of the  radial densities of the Milky Way to emitter galaxy (that galaxy from we are measuring the photons).  This conclusion makes the current version of the LCM quite powerful in relation to both DM theories and other alternative gravitational models.  Furthermore, as can be seen in Fig. 4-6, the rotation curves are predicted at the documented values reported in Table 2, 
without the need for invoking any dark matter. This is because, the relative curvature contributions allow that the luminous mass profile can be extracted from the observed spectra shifts of light. 
\\
It has been  noted by ~\citet{Bot}  that   a credible, empirical alternative to DM   must: 
\begin{itemize}
\item   successfully predict rotation curves with reasonable estimates of the stellar mass-to-light ratios   and gas fractions
\item  make sense within our physics framework
\item  have fewer free parameters or have one universal parameter
\item  predict various other astrophysical observations.
\end{itemize}  
 
While the LCM has not yet been extended to other astrophysical observations, at this early stage of development in proof of concept,   the LCM does the first three very well.  

To predict other astrophysical phenomena, the geometrical construction here in the context of high symmetry must be extended to other more complex  situations. 
In   future work the first places the LCM can be applied is to a statistical constraint to the   Milky Way luminous mass  models and to 
weak lensing.  The LCM is already constructed for testing the first case and in the second, \citep{Narayan} have already phrased the weak lensing problem in similar curvature terms.   Two very interesting next directions include investigations of the coincidence between the dark matter halo density exponent and that of the interpretation of the LCM free parameter $\alpha$ given in Eq.~\ref{correl}, and fitting  the rotation curve of the Milky Way itself.   Since the LCM formalism is based upon the conjecture that the frame-dependent effects of the Milky Way's luminous mass profile are convolved into our observations of spectral shifts, analysis of the Milky Way rotation curve itself will rely upon rephrasing the LCM construction in the case where the observer's frame is imbedded in the emitter's global frame. 
The LCM is potentially falsifiable in any of these cases, and eventually once it has been tested in a larger sample of galaxies, could be in principle extended numerically to an arbitrary metric case of dynamics to include galaxy mergers and galaxy clusters. In the immediate context, the LCM is a robust model which constrains the luminous matter modeling of galaxies directly from spectra, under the assumption of a Milky Way luminous mass profile in a manner that is void of free parameters.

   \section[]{Acknowledgments}\label{sec:fin}
The authors would like to thank R.\,A.\,M.\, Walterbos, V.\,P.\,  Nair,   M.\, Inzunza II,  J.\, Conrad,  V.\, Papavassiliou, T.\, Boyer, P.\, Fisher,   E.\, Bertschinger, I.\, Cisneros, S.\, Nabahe, D.\, Wilmot, T.\, Robertson and R.\ J.\  Moss.   \\
 S.\,  Cisneros has received support from  the MIT Martin Luther King Jr. Fellowship throughout this work, while J.\,A. Formaggio and N.\,A. Oblath are  supported by the United States Department of Energy under Grant No. DE-FG02-06ER- 41420.
 \appendix
\section{LCM Heuristic }
 The most general  form of the  Lorentz transformation is the   exponential mapping:
 \begin{equation}
 \Lambda= e^{\chi}=\sum_{n=0}^{\infty} \frac{\chi^n}{n!}
 \end{equation}
 where   $\chi=-\xi S$   is the  product of the rapidity angle $\xi$ and  the generator of the rotation $S$. The rotation through the   
 angle $\xi$, for a given   action $S$,  defines the relationship between two frames in the hyperbolic space-time of Special Relativity.  In the LCM we take the measured photon frequencies to back-out the underlying luminous mass profile using the Lorentz mapping formalism.
 
   The Doppler-shift formula in Eq.~\ref{eq:dataLorentz}  comes from such a Lorentz transformation; specifically,   
 rotating   a   photon's 4-vector $(\omega,  k_i)$   between two frames,  as related by the rapidity angle $\xi$, for   $\lambda$  the wavelength,  $k_i=2\pi/\lambda_i$, and   the indices of  the   spatial basis $i=1,2,3$.  Written geometrically then, the Lorentz Doppler-shift formula is: 
 \begin{equation}
\frac{v}{c}= \tanh \xi= \frac{e^\xi - e^{-\xi}}{e^\xi + e^{-\xi}}. 
\label{eq:LorentzDefine}
\end{equation}
We can extend  this  formalism to   
to a pair of emitter 
and receiver galaxies by noting that the kernel of this mapping is the ratio of the  received to emitted  frequency: 
  \begin{equation}
  e^\xi(r)=\omega_{\rm{receiver}}(r)/\omega_{\rm{emitter}}(r),
  \label{eq:array}
  \end{equation} 
 and that the diffuse nature of these galactic systems is within the regime of Special Relativity, thus the common use of 
 Newtonian kinematics in the treatment of dark matter analyses.
\subsubsection{ Curved Mapping $v_1$ } 

 The first LCM term, $v_1$,  looks at the gravitational redshift frequencies  $\omega(r)$ (Eq.~\ref{eq:Clone}) of the emitter galaxy with respect to those of the receiver galaxy (eg. Milky Way),  phrased kinematically as  equivalent Doppler-shifts.  The kernel for the curved  $2-$frame map of  emitter to the receiver galaxy,   is identified with   the Schwarzschild   gravitational   redshifts due to the  enclosed  luminous mass as a function of radius.  It has been shown in \citet{Cisn}
  that the addition of Kerr-type effects are nominal in these cases,  within our current observational capabilities, and so use of the Schwarzschild formalism is sufficient to demonstrate the relative curvature effects. 
  
  The  curved  $2-$frame mapping kernel is then: 
 \begin{equation}
e^{\xi_{c}}(r)=\frac{\omega_{mw}(r)}{\omega_{gal}(r)},  
\label{eq:specific}
\end{equation}
  for  $\omega_{gal}(r)$  and $\omega_{mw}(r)$ respectively the gravitational redshifts of the emitter ($gal$) and receiver galaxies ($mw$) as a function of radius (Eq.~\ref{eq:Clone}).

The  curved $2-$frame convolution function $v_1$ has been modified from the  original form reported in \cite{Cisneros:2013vha} and \cite{Cisneros:2014fea}, as we have found  a more robust Lorentz-type convolution to involve only the respective clocks of the two galaxies.   The term $\gamma^{-1}=sech (\xi_c)=d\tau/dt$, familiar from 
Special Relativistic treatments of mass and time dilation, acts to  relate the  clocks between different frames using shifted photon frequencies as the measure of clock time in the manifold.  The more robust 
   $2-$frame curvature map $v_1$  is then:
\begin{equation}
\frac{v_{1}}{c}=\left(\frac{ 2}{e^{\xi_{c}}+ e^{-\xi_{c}}} -1\right).
\label{eq:prime1}
\end{equation}
All quantities are     functions of radius except for the vacuum light speed.   

\subsubsection{Curved to Flat  Mapping  $v_2$ } 
 The second LCM term, $v_2$,  transforms between the curved $2-$frame  (kernel     Eq.~\ref{eq:specific}) to the flat $2-$frame.  Since all physics measurements are made in our local flat frames, it is necessary to now transform from the previous curved $2-$frame map to the flat $2-$frame.   The reported Keplerian rotation curves $v_{lum}(r)$,  which are frequently reported with the observed ``flat'' rotation curves to show discrepancies at large radius,  are  our best estimates of these flat-frames and so will be used to define the flat $2-$frame.
 
The shifted frequencies  $\omega_{l}(r)$  expected for $v_{lum}(r)$ are defined by the relation:
     \begin{equation}
 \frac{v_{lum}(r)}{c}=
 \frac{
 \frac{\omega_{l}(r) }{\omega_{o}}
-  \frac{\omega_{o} }{\omega_{l}(r)}}{ 
  \frac{\omega_{l}(r)}{\omega_{o}}
+  \frac{\omega_{o} }{\omega_{l}(r)}} .
 \label{eq:MPflat}
\end{equation} 
Consistent with   Eq.~\ref{eq:array}, the   flat 2-frame mapping  kernel  is:
 \begin{equation}
  e^{\xi_{f}}(r) =   \frac{\omega_{l} (r)}{\omega_{o}}.
    \label{eq:flatflat}
     \end{equation}
Since the $v_2$  mapping involves four frames ($2-$frame onto $2-$frame)  it is algebraically convenient to write it as: 
\begin{equation}
 (e^{ \xi_2} )^2= \frac{e^{\xi_{f}}}{ e^{\xi_{c}}  },
 \label{eq:FCFtwo}
\end{equation} 
 such that the final term is: 
 \begin{equation}
\frac{v_2 }{c}=  \frac{e^{ 2\xi_2  }+1}{e^{ 2\xi_2 } - 1}, 
\label{eq:hyperbolico}
\end{equation}  
 where   all quantities are     functions of radius except for the vacuum speed of light and the characteristic frequency $\omega_o$.  We find that this term is most robust when normalized by it's value at large radius, as representative of the value as $r \to \infty$. \\
 It is important to note that this is essentially an enforced reverse boost,
since in  Special Relativity Lorentz boosts are always defined as   positive rotations  away from the rest frame (vertical axis in the light cone   compare  to  Eq.~\ref{eq:LorentzDefine}),    and  here we want to transform back \emph{to} the rest frame. 
 
  In our quest for every refined measurements of the light distribution from distant objects, the LCM is an empirical construction which predicts rotation curves exceedingly well. In so far as it allows an observationally based and expedient manner in which to  discriminate  model based predictions, the LCM is a constraint to  population synthesis modeling which can   expand of our analysis capability. 
 
 \bibliography{LCM}{}
\bibliographystyle{mn2e}
 \bsp
 
\label{lastpage}
\end{document}